\documentclass[a4paper,11pt]{article}
\usepackage{amsmath,amssymb,amsthm}
\usepackage{dsfont}
\usepackage{graphicx}
\usepackage[british]{babel}
\usepackage[utf8]{inputenc}
\usepackage{xcolor}
\usepackage[colorlinks=true,linkcolor=blue,citecolor=red]{hyperref}
\usepackage{authblk}
\usepackage{blindtext}
\usepackage{epstopdf}
\usepackage{cite}
\usepackage{caption,subcaption}

\newtheorem{remark}{Remark}

\newcommand{\R}{\mathbb{R}}
\newcommand{\be}{\begin{equation}}
\newcommand{\ee}{\end{equation}}
\newcommand{\fer}[1]{(\ref{#1})}

\def\CI {\mathcal{I}}
\def\CU {\mathcal U}
\def\z {\mathbf z}

\newenvironment{equations}{\equation\aligned}{\endaligned\endequation}

\begin{document}


%
%

\title{A data-driven epidemic model with social structure for understanding the COVID-19 infection on a heavily affected Italian Province}


\author[2]{M. Zanella\thanks{\tt mattia.zanella@unipv.it}}
\author[4]{C. Bardelli\thanks{\tt chiara.bardelli01@universitadipavia.it}}
\author[1]{G. Dimarco\thanks{\tt giacomo.dimarco@unife.it}}
\author[5]{S. Deandrea\thanks{\tt Silvia\_Deandrea@ats-pavia.it}}
\author[5]{P. Perotti\thanks{\tt Pietro\_Perotti@ats-pavia.it}}
\author[5]{M. Azzi\thanks{\tt Mara\_Azzi@ats-pavia.it}}
\author[4]{S. Figini\thanks{\tt silvia.figini@unipv.it}}
\author[2,3]{G. Toscani\thanks{\tt giuseppe.toscani@unipv.it}}

\affil[1]{\normalsize
        Department of Mathematics and Computer Science, University of Ferrara, Italy}
\affil[2]{\normalsize
         Department of Mathematics "F. Casorati", University of Pavia, Italy}
\affil[3]{\normalsize Institute for Applied Mathematics and Information Technologies (IMATI), Italy}
\affil[4]{\normalsize Department of Political and Social Sciences, University of        Pavia, Italy}  
\affil[5]{\normalsize Health Protection Agency (ATS), Viale Indipendenza, Pavia, Italy}

%
%
%
%
%
%
%
%
%
%
%

\date{}
\maketitle


\begin{abstract}
In this work, using a detailed dataset furnished by National Health Authorities concerning the Province of Pavia (Lombardy, Italy), we propose to determine the essential features of the ongoing COVID-19 pandemic in term of contact dynamics. Our contribution is devoted to provide a possible planning of the needs of medical infrastructures in the Pavia Province and to suggest different scenarios about the vaccination campaign which possibly help in reducing the fatalities and/or reducing the number of infected in the population. The proposed research combines a new mathematical description of the spread of an infectious diseases which takes into account both age and average daily social contacts with a detailed analysis of the dataset of all traced infected individuals in the Province of Pavia. These information are used to develop a data-driven model in which calibration and feeding of the model are extensively used. The epidemiological evolution is obtained by relying on an approach based on \textcolor{black}{statistical} mechanics. This leads to study the evolution over time of a system of probability distributions characterizing the age and social contacts of the population. One of the main outcomes shows that, as expected, the spread of the disease is closely related to the mean number of contacts of individuals. The model permits to forecast thanks to an uncertainty quantification approach and in the short time horizon, the average number and the confidence bands of expected hospitalized classified by age and to test different options for an effective vaccination campaign with age-decreasing priority.
\end{abstract}



\tableofcontents

\section{Introduction}
The recent spreading of the COVID-19 epidemic led the governments of most countries in the world to introduce restrictions such as social distancing and lockdown policies and Italy has not been exempted from these actions.\footnote{\texttt{Chronology of main steps and legal acts taken by the Italian Government for the containment of the COVID-19 epidemiological emergency: http://www.protezionecivile.gov.it/documents/20182/1227694/Summary+of+\\ measures+taken+against+the+spread+of+C-19/c16459ad-4e52-4e90-90f3-\\c6a2b30c17eb}} These \emph{non-pharmaceutical} intervention measures, however, entail significant social and economic costs,\cite{Ash,DPTZ} and it is not thinkable that policy makers will be able to maintain them for more than a limited period of time. Starting with its appearance in 2019, and pending the development of a vaccine, the main strategy to slow down the spread of the epidemic as much as possible, thus avoiding the collapse of the health systems, has been indeed the introduction of lockdown measures. \textcolor{black}{An important feature of these measures} is related to the fact that new data are collected and characteristics of the pandemic are unveiled every day, and these could deserve to be considered to improve the response. In this context, predictive mathematical models are of fundamental importance to understand the course of the epidemic and to plan \textcolor{black}{measures to mitigate fatalities}, hospitalizations and the social impact of COVID-19 and of the lockdown policies.\cite{Est}

Among the most known models for describing epidemic dynamics is the so-called SIR model,\cite{Anderson,BCF,DH,HWH00,SIR} which furnishes the number of individuals through three mutually exclusive compartments: Susceptible, Infected and Recovered. More complex models can be certainly derived and can describe more in details the epidemic such as the SIAR model,\cite{Chisholm} in which Asymptomatic are taken into account or the SEIAR one,\cite{Gatto} in which Exposed individuals are also considered. However, one main limitation in the direction of enriching the model is the difficulty associated to the calibration of the parameters needed \textcolor{black}{to bridge such mathematical descriptions} to the real world application which in the specific case of COVID-19 remains of paramount importance. In many cases, data sets are not even at disposal and the setting of complex models, involving many parameters, is done based on an empirical analysis. Another fundamental aspect regards the fact that data are affected by errors and a way to measure this uncertainty should be set into place.

For the COVID-19 pandemic, several models have been already developed and used for forecasting \cite{APZ,Ana,Bell,Britton,DPeTZ,Gatto,Bruno,Jiang,PS}.  Many of these models, however, are based on the hypothesis of homogeneity of the population inside a compartment. For example, in Ref.  \cite{Ana}, a SIRD model is used and fitted with data, while in Ref.  \cite{Gatto} a quite comprehensive model in terms of compartments have been considered but without distinguishing about social attitudes or age while in Ref.  \cite{Bruno}, the authors discriminates between detected and undetected cases and between different severity of illness. In all the cited cases and in many others, homogeneity of the population inside a compartment is hypothesized, i.e. nor distinction of ages nor distinction about the average number of contacts is done. This simplifying choices have the evident advantage to deal with a model depending on a limited number of parameters, which in principle can be accurately fitted starting from reality in a relatively simple way. As recently discussed in Ref.  \cite{DolT} for instance, one of the advantages of simple models is that the impact of the variation of a single parameter on various qualitative properties can be easily studied. On the other hand, there is a certain risk in validating outcomes because of a too marked simplification of the representation of a complex system. On the top of that, it is always present the additional difficulty that some parameters cannot be easily quantifiable from the available statistical data or data cannot be available and when available affected by uncertainty. In Ref.  \cite{Britton}, a step forward about the heterogeneity of the population in terms of contact rate has been considered, the same holds true for Ref.  \cite{APZ} where the impact of the different ages were analyzed and uncertainty taken into account.

\textcolor{black}{Here, we propose a new multiscale approach which is based on two main ingredients that play a leading role in the dynamics: the dependence on the age of the evolution of the disease and the dependence of the grow rate of the epidemic on the average behavior of the population in terms of number of daily contacts \cite{age,Moss}.
This  approach is coherent with the recent study of a multidisciplinary modeling of the pandemic developed in Ref.  \cite{Bell}, which outlines the importance of taking simultaneously into account the different scales (from the  microscopic to the mesoscopic one) which are involved in the epidemic, and suggests, among other  factors to be considered in the epidemic spreading, the key role of the heterogeneity of the society. }
These two aspects produce significant changes in the time evolution of the epidemic well-supported by the increasing evidence of the contribution to the spread of the infection caused by asymptomatic patients \cite{Gaeta,LTB}. In fact, the speed at which the epidemic spreads suggests that transmission from unidentified (because asymptomatic) cases plays a substantial role in the exponential increase in the number of infected patients. The main consequence is that, while mathematical models in the context of COVID-19 have demonstrated their usefulness in simulating transmission trends, risk assessment, epidemic development, and the effects of isolation and quarantine,\cite{Jiang,LPR,McC,Mizu,Ril} missing the effect of asymptomatic population introduces errors into prediction models \cite{Gaeta}.  On the other hand, it is an established fact that the mortality from COVID-19 among infected population is increasing with age,\cite{age} and this causes critical situations in the Health System. \textcolor{black}{Finally, the leading role of the social contacts is nowadays clear and therefore it should be considered in the construction of realistic models}. An interesting approach to the quantification of the effects of limited social interactions in the spreading of COVID-19 epidemic has been recently proposed in Ref.  \cite{DPeTZ}, where the classical SIR model has been coupled with a kinetic description of a multi-agent system \cite{PT13}.  In this approach, the individuals are characterized by the distribution over time of the number of their daily social contacts. There, the elementary interaction charactering at a kinetic level the formation of the \textcolor{black}{emerging} distribution of social contacts has been modeled in agreement with recent studies of the scientific community aimed at estimating the distribution of contacts between individuals. This was recognized as a relevant cause of the potential pathogen transmission of infectious diseases (cf. Ref.  \cite{Bell,Plos,Fuma,Moss} and the references therein). \textcolor{black}{The introduction of a detailed interaction law based on the number of social contacts leads, at the macroscopic observable level of description, to a consistent saturated incidence rate depending on the evolution of the infection itself. The resulting model is therefore derived from the microscopic behavior of interacting agents and not heuristically postulated.}
 Kinetic-type descriptions have been often fruitfully developed to gain an insight on system composed by many agents, without pretending to review to very vast literature on this topic we mention here Ref.  \cite{Albi,BD,GT19,PT13,To4}.

\textcolor{black}{The above discussed characteristics suggest that, to follow the time evolution of COVID-19 infection and to control it at best with non-pharmaceutical interventions, robust mathematical models need to take into account a number of new variables, which include at least the age, the number of social interactions and the compartment describing the  asymptomatic population.} A fundamental aspect regards also the quantification of the uncertainties since many characteristics of the epidemic cannot be known with precision. On the top of the choices done for the mathematical models, one has always to take into account that dataset are strictly necessary to produce reliable models and thus, introducing additional parameters which cannot be estimated and extrapolated from data is always a risk which in principle has to be avoided when possible. Compared to the model proposed in Ref.  \cite{DPeTZ} here we enrich the description by considering the effects of age on the evolution of the disease and the effects of vaccination on reducing the number of fatalities and avoiding picks of infected which possibly effect the health system. The dynamics are also enriched by introducing uncertain variables into the system and their relative estimation based on the available clinical data.

In order to calibrate the proposed model we followed a two step approach. In details, we first determine relevant epidemiological parameters before the adoption of strict governmental limitations to contain the second wave of the pandemic, then the relative reduction of the contact rate has been determined using a dynamic approach where we match our model with the dataset using the determined incidence rate through the number of asymptomatic and the average contact rates divided per ages. Subsequently,  we concentrate on the possibility to forecast the epidemic trends and we discuss vaccination campaign under different strategies in terms of expectation and relative confidence bands.

The rest of the paper is organized as follows. In Section \ref{sec:stat} we describe the dataset and we perform an exploratory analysis. In Section~\ref{sec:model} we introduce a system of SIAR-type kinetic equations combining the dynamics of social contacts with the spread of the infectious disease in a system of interacting individuals belonging to the four classes of susceptibles, infected, asymptomatic and recovered. The distribution functions of individuals in the classes depend on both the age and the social contact variables. In addition, the percentage of immunized population over time as a result of vaccination by decreasing age groups is included in the model. In this way, the possibility to have at disposal a system in which the age of individuals is an independent variable, allows to recover in a precise way both the evolution of the pressure in time on the health system, which is recognized to depend on age, and the effects in time of a vaccination campaign made according to the rule of age-decreasing priority. Some qualitative examples of the model capabilities are also shown in this part. In Section \ref{sec:calibration}, we determine all parameters intervening in the model through the use of data of the province of Pavia (Lombardy Region, Italy). \textcolor{black}{In the same section, we show forecasting results, validating them with experimental data, and we analyze  the impact of several vaccination strategies}. \textcolor{black}{Finally, in Section \ref{sec:conc} we draw some conclusions and possible future developments. }


\section{Dataset and statistical analysis}\label{sec:stat}
A statistical analysis has been performed on the dataset provided by \href{https://www.ats-pavia.it/}{ATS} (Health Protection Agency) of Pavia, Italy, to identify significant correlations and dependencies between the evolution of the COVID-19 and the age of the infected people in terms of the relevant parameters of the epidemic. This analysis is a fundamental step for the construction of our data-driven model (S-SIAR, Social SIAR, model in the following). \textcolor{black}{ In particular, the recovery rates divided per age group as well as the number of hospitalized patients and deaths which can be extrapolated from data,  will be shown next and successively integrated in the model. }

\begin{figure}
    \centering 
    \begin{subfigure}{0.8\textwidth}
        \includegraphics[width=\linewidth]{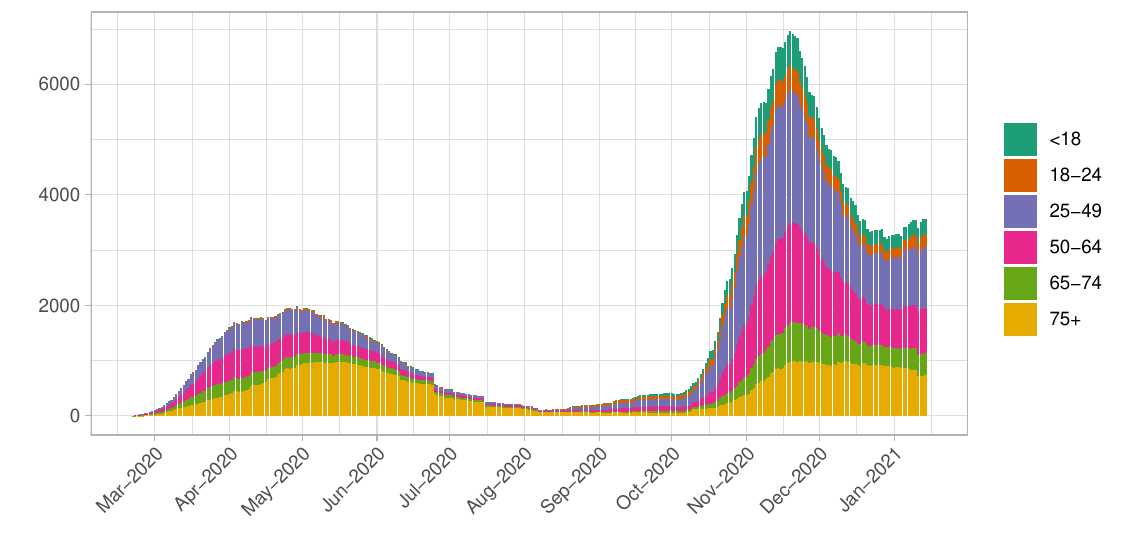}
        \caption{Absolute number of confirmed COVID-19 cases in the Pavia province divided per class of age. Data start on February 2020 and end in January 2021.}
        \label{fig:cumul_inf}
    \end{subfigure}
    \medskip 
    \begin{subfigure}{0.8\textwidth}
        \includegraphics[width=\linewidth]{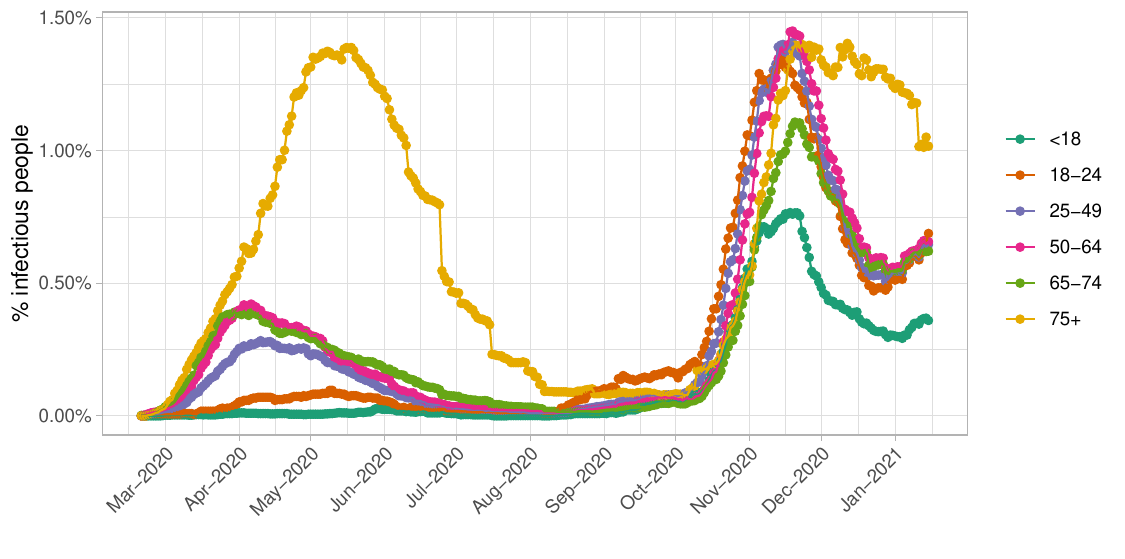}
        \caption{Percentage of confirmed COVID-19 cases over the total number of individuals for each age group.}
        \label{fig:evolution_inf_age_perc}
    \end{subfigure}
    \caption{Epidemic curves for each age group in the Pavia province. Data start on February 2020 and end in January 2021.}
    \label{fig:hist_tvc}
\end{figure}

The dataset includes all the confirmed COVID-19 cases of Pavia from February 20, 2020, to January 15, 2021. Figure \ref{fig:cumul_inf} shows the epidemic curve of traced infected individuals for each age group in the considered period. As it appears clear from Figures \ref{fig:cumul_inf} and \ref{fig:evolution_inf_age_perc}, despite for the majority of the population the infection shows two distinct peaks, the epidemic curves are characterized by different shapes, both in terms of absolute numbers and percentages. The first peak reveals \textcolor{black}{a high} number of over 75 individuals infected, while the second peak is more homogeneous among the considered age groups. This effect is probably due to the fact that during the so-called first wave, the tracing methodologies were not so effective and many asymptomatic or pauci-symptomatic individuals have not been identified and more likely this type of infected belong to younger age classes.
The above dataset has been used to understand and extrapolate the so-called Time of Viral Clearance (TVC) for each age category. This quantity is estimated both for recovered patients and deceased individuals from the difference between the first positive swab test and the first negative swab test (or the death date). The statistical analysis is performed on TVC considering only the months of October, November and December since this period includes the most complete and accurate data provided by ATS of Pavia and it is considered the most reliable one by ATS. All the patients with a TVC less than 5 days have been excluded from the statistical analysis, as they can be the result of erroneous data entries. Figure \ref{fig:hist_tvc} shows the histograms of TVC (in days) related to the six age groups considered namely: 0-17, 18-24, 25-49, 50-64, 65-74 and over 75. The Table \ref{tab:age_param} reports on the left the corresponding parameters which describe the estimated distributions. The right table \ref{tab:age_param} shows the composition of each studied group. The support of each histogram has been computed excluding all the outliers detected by the interquartile range ($IQR$) rule: all the values greater than the 75th percentile $+ 1.5*IQR$ are considered outliers. For each histogram a Beta density function has been fitted using the method of the Maximum Likelihood Estimation to estimate the parameters $\alpha$ and $\beta$. The distribution then reads as
\be\label{beta} 
f_{\alpha,\beta}(x)=\frac{x^{\alpha-1}(1-x)^{\beta-1}}{\int_{0}^{1}x^{\alpha-1}(1-x)^{\beta-1}dx},
\ee

where $x$ is measured in days. The Beta density function is properly scaled to fit the bounded supports for each age group reported in Table \ref{tab:age_param}.

\begin{figure}
    \centering 
    \begin{subfigure}{0.31\textwidth}
        \includegraphics[width=\linewidth]{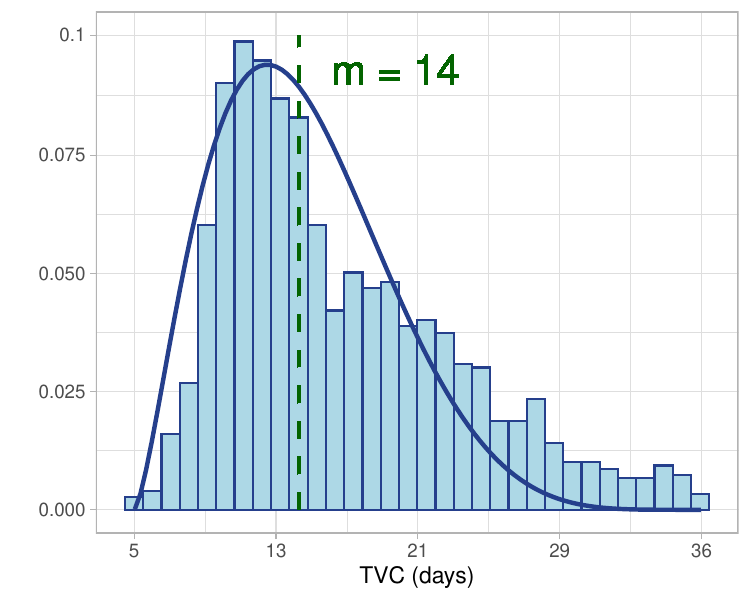}
        \caption{Group 0-17}
        \label{fig:hist_tvc_0_17}
    \end{subfigure} 
    \begin{subfigure}{0.31\textwidth}
        \includegraphics[width=\linewidth]{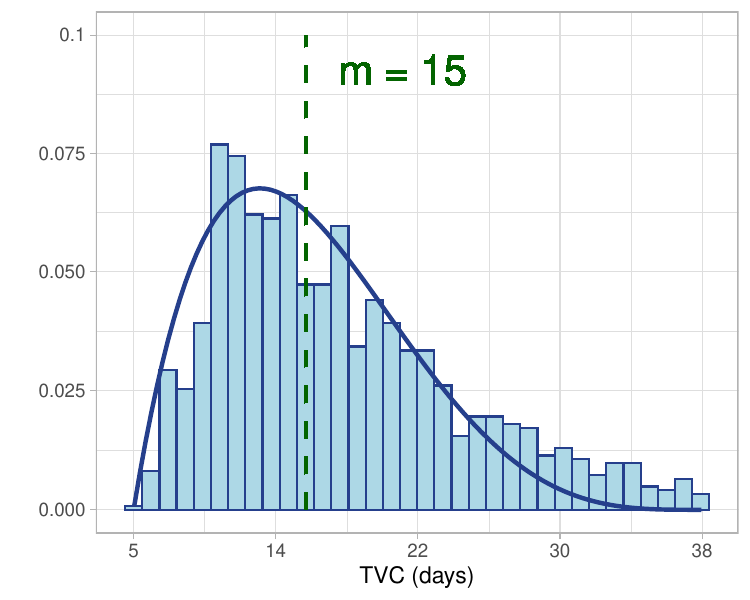}
        \caption{Group 18-24}
        \label{fig:hist_tvc_18_24}
    \end{subfigure} 
    \begin{subfigure}{0.31\textwidth}
        \includegraphics[width=\linewidth]{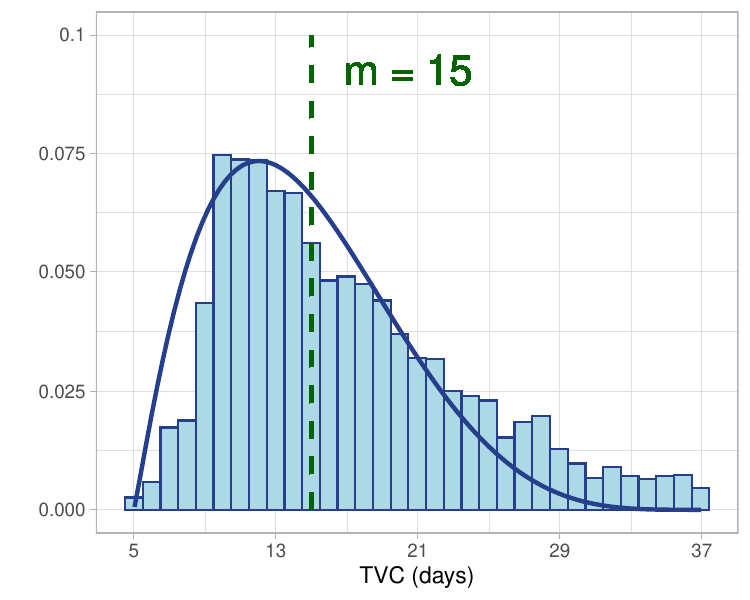} 
        \caption{Group 25-49}
        \label{fig:hist_tvc_25_49}
    \end{subfigure}
    
    \medskip
    \begin{subfigure}{0.31\textwidth}
        \includegraphics[width=\linewidth]{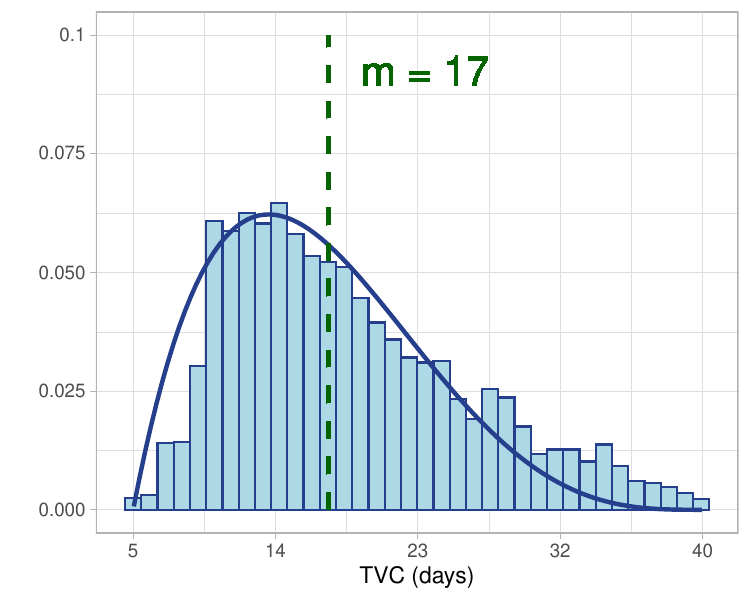}
        \caption{Group 50-64}
        \label{fig:hist_tvc_50_64}
    \end{subfigure} 
    \begin{subfigure}{0.31\textwidth}
        \includegraphics[width=\linewidth]{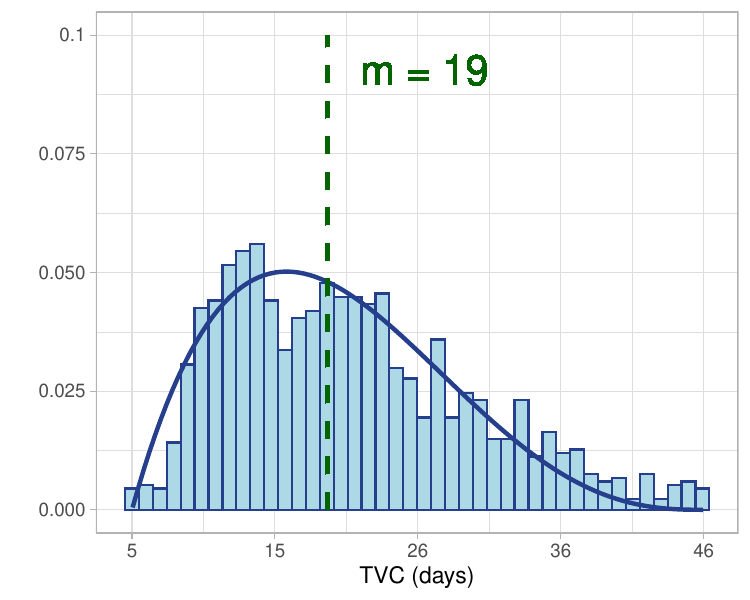}
        \caption{Group 65-74}
        \label{fig:hist_tvc_65_74}
    \end{subfigure} 
    \begin{subfigure}{0.31\textwidth}
        \includegraphics[width=\linewidth]{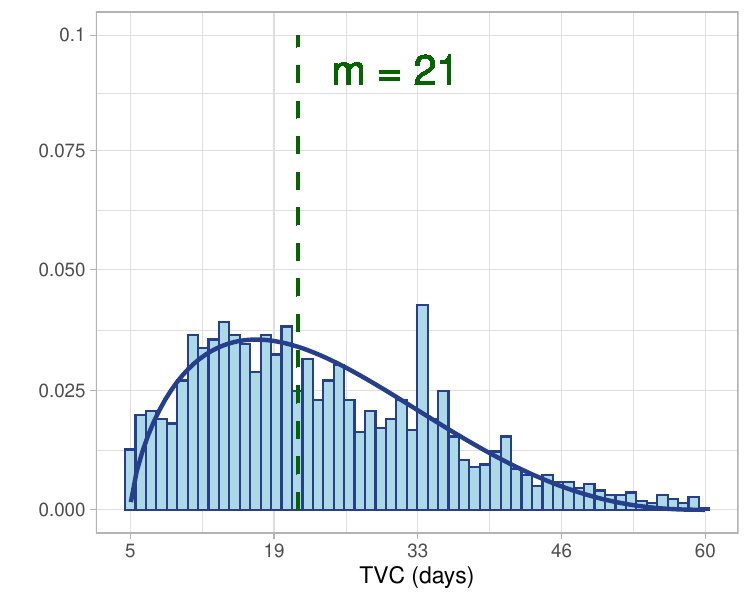}
        \caption{Group over 75}
        \label{fig:hist_tvc_75}
    \end{subfigure}
    \caption{Histograms of TVC for each age group and beta density functions fitted on data. The green dashed line represents the median TVC computed for each group age.}
    \label{fig:hist_tvc}
\end{figure}
\begin{figure}
    \begin{center}
        \includegraphics[scale = 0.7]{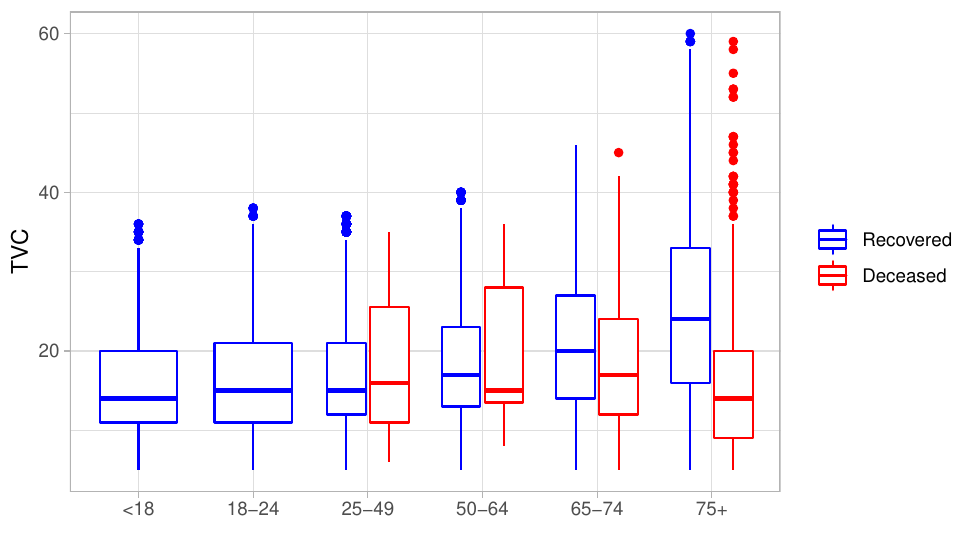}
        \caption{TVC distribution divided for recovered and deceased individuals for each age group.}
        \label{fig:box_tvc_dec}
    \end{center}
\end{figure}

Table \ref{tab:age_param} shows that the median value of TVC increases with the age of patients. The shape of the Beta density function fitted on data reveals an high peak around lower values for younger patients (under 18 years old) and a more uniform behaviour for older patients (over 75 years old); changing the support from $[5,36]$ to $[5,60]$  shows that young individuals recover quickly from the disease, while COVID-19 persists longer in older patients. To deeper study the distribution of TVC comparing recovered and deceased people, Figure \ref{fig:box_tvc_dec} presents the results for each age group. For younger classes of age, there is no significant differences between distributions of recovered and deceased people. On the other hand, the oldest age groups show a clear gap with shorter TVC for deceased individuals, suggesting that the most severe cases of COVID-19 in older patients get worse very quickly.
\begin{table}
    \centering
    \begin{tabular}{lcccc}
        \hline\noalign{\smallskip}
        Age group & Support & $\alpha$ & $\beta$ & Median  \\
        \hline
        0-17  & $[5,36]$ & 2.5 & 5.9 & 14 \\
        18-24 & $[5,38]$ & 2.0 & 4.5 & 15 \\
        25-49 & $[5,37]$ & 2.1 & 4.9 & 15 \\
        50-64 & $[5,40]$ & 2.0 & 4.2 & 17 \\
        65-74 & $[5,46]$ & 2.0 & 3.7 & 19 \\
        over 75 & $[5,60]$ & 1.7 & 3.5 & 21 \\
        \noalign{\smallskip}\hline
    \end{tabular}\qquad 
    \begin{tabular}{lcc}
    \hline\noalign{\smallskip}
    Age group & Group size & Percentage  \\
    \hline
    0-17  & 80425 & 14.72 \\
    18-24 & 34201 & 6.26  \\
    25-49 & 172596 & 31.59  \\
    50-64 & 124835 & 22.85  \\
    65-74 & 63696 & 11.65  \\
    over 75 & 70762 & 12.95  \\
    \noalign{\smallskip}\hline
\end{tabular} 
    \caption{Left: distributions parameters for each age group using the Maximum Likelihood Estimation on a beta distribution. Right: number of individuals per class and relative percentages.}
    \label{tab:age_param}
\end{table}

\begin{figure}
    \begin{center}
        \includegraphics[scale = 0.6]{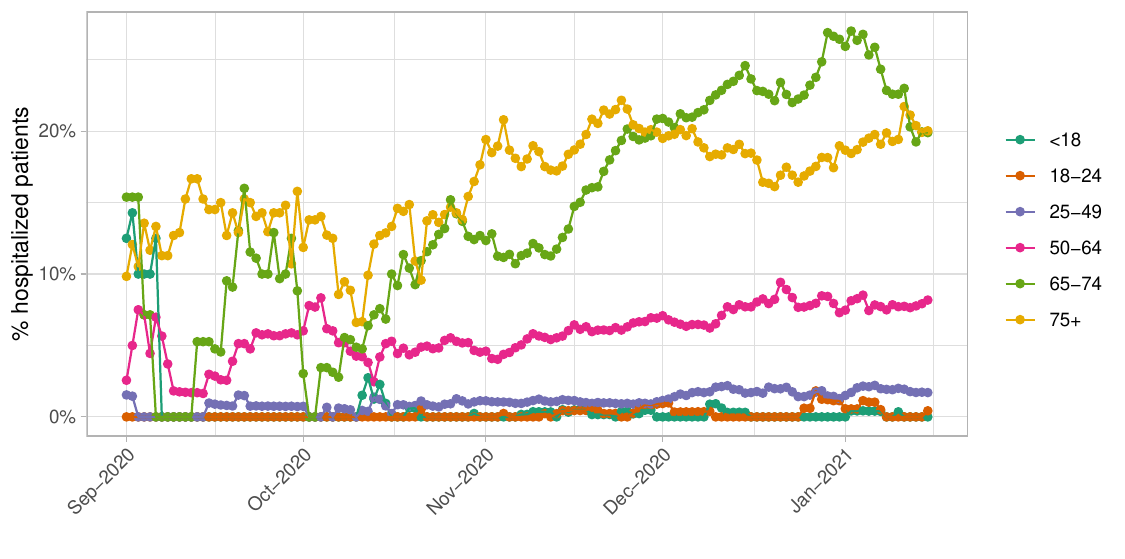}
        \caption{Daily percentage of hospitalized patients over the total number of infected people per each age group in the Pavia province.}
        \label{fig:hosp_patients_perc}
    \end{center}
\end{figure}

The prediction of the number of infected people given by the model can be also exploited to estimate the number of hospitalized patients as done in Ref.  \cite{Zanella} through a simpler model not involving the age of the population. This is a crucial aspect for the achievement of an optimal organization of the healthcare system in case of emergency. The number of hospitalized individuals can be directly derived from the number of positive COVID-19 cases computing the percentage of hospitalization from data at hand since this feature is not expected to change as a function of the time evolution of the pandemic. For this reason, in our model, this quantity is extrapolated through a post-processing step, applying the derived hospitalization percentages to the prediction of the number of infectious individuals. Figure \ref{fig:hosp_patients_perc} shows this percentage for each age group with a particular focus on the period September 1, 2020 and January 15, 2021. The younger age groups (from 0 to 24 years old) present a very low percentage, almost null, for the entire period. The percentage of the group 25-49 is slowly growing with a mean value of 1.1\%, as the percentage of 50-64 which is higher but with a similar small growth around the mean value of 6\%. The older age groups are subjected to considerable fluctuations which make the estimate of the percentage of hospitalization more difficult with a mean around 15\%. One can infer that, being the percentage of hospitalizations almost constant per age class and independent on the period of time (there are no evidence of more dangerous mutations), the total number of infected is larger than the ones traced by the health care system and this becomes especially true starting from November 2020. 

\section{The age-structured S-SIAR model}\label{sec:model}

The goal of this section is to introduce a new model to describe the spreading of an infectious disease under a dependence of the contagiousness parameters on the individual number of social contacts of the individuals, identified in terms of their age. Unlike the classical SIR models,\cite{HWH00}  and following the modeling assumptions of Ref.   \cite{LTB,Gaeta}, the entire population is divided into four classes: susceptible (S), infected (I), asymptomatic (A) and  recovered people (R). This last class includes formerly symptomatic removed and formerly asymptomatic removed (mostly passing unnoticed through the infection).

The objective is to understand the joint effects of age and social contacts on the dynamics. Therefore, we will consider in the sequel a system of agents characterized by the age variable $x$, measured in years, with $x \in \CI = (0,100)$, and the variable $v \ge0$, representing the number of daily contacts of an individual. With respect to the social contact variable $v$, agents in the system are considered indistinguishable \cite{PT13}. This means that, for any given value of the age variable $x$, the state of a person in each class at any instant of time $t\ge 0$ is completely characterized by the number of contacts $v \ge0$. The role of age is here similar to the role of heterogeneity in the disease parameters such as the personal susceptibility to a given disease as suggested in Ref.  \cite{Diek,Novo,Van}. 

While $v$ is a natural positive number at the individual level, without loss of generality we will consider $v$ in the rest of the paper to be a nonnegative real number, $v\in \mathbb{R}_+$, at the population level. We denote then by $f_S(x,v, t)$, $f_I(x,v,t)$, $f_A(x,v,t)$ and $f_R(x,v,t)$, the probability distributions at time $t > 0$ of the number of social contacts of the population of age $x$ of susceptible, infected, asymptomatic and recovered individuals, respectively. The chosen notation tends to highlight the analogies between the present situation and the classical kinetic theory of rarefied gases. This theory is helpful to justify the forthcoming modeling assumptions by resorting to the well-consolidated framework of the passage from the mesoscopic description in terms of the classical Boltzmann equation to the fluid-dynamics description based on fluid equations (either Euler or Navier-Stokes). These aspects even if necessary from the theoretical point of view will move too far away the discussion with respect to the core of the present work. For this reason, we postpone the details to the Appendix \ref{sec:kinetic}. Resorting to the analogy with the kinetic theory, we assign to the age variable $x$ the role of the position variable, and to the contact variable $v$ the role of the velocity variable. Hence, the \emph{local} densities $\rho_J(x,t)$, $J\in \{S,I,A,R\}$, of the population of the various classes in terms of the age $x\in \CI$ at time $t \ge 0$ are obtained by integration with respect to the contacts variable $v$ of the distribution functions $f_J(x,v, t)$, $J\in \{S,I,A,R\}$
 \be\label{densi}
 \rho_J(x,t) = \int_{\R_+} f_J(x,v,t) \, dv,
 \ee
Analogously, we can define other local moments like the mean number of contacts of each class $c_J(x,t) $, $J\in \{S,I,A,R\}$, of the population of age $x\in \CI$ at time  $t\ge 0$ by 
\be\label{mean-number}
 c_J(x,t) = \int_{\R_+}v f_J(x,v,t) \, dv,
 \ee
from which we can compute the \emph{local} mean density of contacts $m_J(x,t)$ $J\in \{S,I,A,R\}$, of the population of age $x\in \CI$ at time  $t$
\be\label{mean-densi}
 m_J(x,t) = \frac 1{ \rho_J(x,t)}\int_{\R_+}v f_J(x,v,t) \, dv.
 \ee
The local density of the whole population of age $x$ is then recovered as the sum of the  $\rho_J(x,t)$
\[
\rho(x,t)= \sum_{J\in S,I,A,R}\rho_J(x,t).
\]
As outlined in the Introduction, since we do not consider for simplicity of presentation disease related mortality, we can fix the total density to be the conserved probability density for all times $t \ge 0$
\[
\int_{\CI} \rho(x,t)\,dx = 1. 
\]
As a consequence, the quantities 
\begin{equation}\label{mass}
J(t)=\int_{\CI}\rho_J(x,t)\,dx,\quad J \in \{S,I,A,R\}  
\end{equation}
denote the fractions of susceptible, infected, asymptomatic and recovered at time $t \ge 0$, respectively. Let us observe that considering the fatalities in the modeling would not change the \textcolor{black}{dynamics} and the results in terms of number of infected will remain the same. For this reason, we decided to extrapolate the number of deaths through a postprocessing procedure starting from the recovered individuals and using the dataset information about the mortality. 

In what follows, we assume that the evolution in time of the distribution functions  $f_J(x,v,t)$, $J\in \{S,I,A,R\}$ is built according to the classical epidemiological models containing in particular the asymptomatic individuals.  Further, we include in the model the possibility to move directly from the class of susceptible to the class of recovered in presence of a vaccination campaign identified by the parameter $\chi(x,v,t)$ which in general may depend on age, number of contact (work status of the individuals) and time (possibility to have access to vaccines). The kinetic model then follows combining the epidemic process with the contact dynamics in which the various classes in the model could act differently in the social process. This gives the system 
\begin{equations}\label{siar-gamma}
&\frac{\partial }{\partial t}f_S(x,v,t) \\
&\qquad= -K(f_S,f_I+f_A)(x,v,t) - \chi(x,v,t) +  \frac 1\tau Q_S(f_S)(x,v,t)
\\
&\frac{\partial }{\partial t} f_I(x,v,t) \\
&\qquad= \xi(x) K(f_S,f_I + f_A)(x,v,t)  - \gamma_I(x) f_I(x,v,t) + \frac 1\tau Q_I(f_I)(x,v,t)
\\
&\frac{\partial }{\partial t} f_A(x,v,t) \\
&\qquad= (1-\xi(x))K(f_S,f_I+ f_A)(x,v,t)  - \gamma_A(x) f_A(x,v,t) + \frac 1\tau Q_A(f_A)(x,v,t)
\\
&\frac{\partial }{\partial t} f_R(x,v,t) \\
 &\qquad=  \gamma_I(x)  f_I(x,v,t) + \gamma_A(x)  f_A(x,v,t)  +\chi(x,v,t) + \frac 1\tau Q_R(f_R)(x,v,t), 
\end{equations}
where we suppose that $\chi(x,v,t)$ is such that
\be\label{vac}
0\leq \chi(x,v,t)\leq f_S(x,v,t)
\ee
which means that at a given instant of time $t$ the number of susceptible individuals belonging to a certain class decrease less than exponentially fast as a consequence of vaccination. In fact, this term quantifies the amount of susceptible individuals which are immunized in time, and consequently move directly from the class of susceptible to the class of recovered. This has been chosen depending to the age variable $x$ to describe the situation in which the vaccination campaign is operated by acting on different class of ages. This coefficient is also assumed to vanish for $x \le \tilde x$, thus excluding people with an age below $\tilde x$ from vaccination. For the sake of generality, we consider the possibility that the intensity of the administration of the vaccine could depend on time and on the number of contact $v$ as already discussed. The function
 $K(f_S,f_I+f_A)$ is the local incidence rate. One simple possibility is to assume it of the form \cite{DPeTZ}
 \be\label{inci}
 \begin{split}
& K(f_S,f_I+ f_A)(x,v,t) \\
 &\qquad=f_S(x,v,t) \int_{\CU(x)} \int_{\R^+} \kappa(v,w)(f_I(y,w,t)+ f_A(y,w,t)) \,dw\, dy.
 \end{split}\ee
 This function governs the transmission of the infection in terms of the social contacts between susceptible individuals of a certain age $x$ having a number $v$ of contacts with the population of individuals which can transmit the disease (infected and asymptomatic individuals). In \fer{inci}, $\CU(x) \subseteq \CI$ denotes the part of the population of infected and asymptomatic which can transmit the disease to a person of age $x$. Then, the age-dependent percentage of individuals which pass from the class of susceptible to the class of infected is \textcolor{black}{defined through to the age-dependent} function  $0< \xi(x)<1$. Note that in \fer{inci} it is assumed that both infected and asymptomatic individuals can transmit the disease at the same rate. Clearly, as proposed in Ref.  \cite{LTB}, one could distinguish the rates of transmission, at the price of introducing a further parameter in front of one of the two classes. Let observe however, that considering the same transmission rate for infected and asymptomatic does not imply that they infect other individuals at the same rate since this also depends on the average number of contact which can be and will be considered different for the two compartments. The contact function $\kappa(v,w)$ in \fer{inci} is a nonnegative function growing with respect to the number of contacts $w$ of the population of infected and asymptomatic, such that $\kappa(v,w=0) = 0$. A leading example is obtained by choosing a rank~1 matrix of type
 \[
 \kappa(v,w) = \beta\, v^\alpha w^\alpha,
 \]
 where $\alpha, \beta$ are positive constants, that is by taking the incidence rate directly proportional to the product of the number of contacts between susceptible and both infected and asymptomatic people. Following Ref.  \cite{DPeTZ}, we will choose $\alpha=1$, which implies that the incidence rate takes the form
  \be\label{simple}
 K(f_S,f_I+f_A)(x,v,t) = \beta \, v f_S(x,v, t)\int_{\CU(x)} \left(c_I(y,t) +c_A(y,t)\right)\, dy,
 \ee
\textcolor{black}{we refer to Ref.  \cite{DPeTZ} for a more general case with arbitrary $\alpha>0$.}  Coming back to equation \fer{siar-gamma}, the functions  $0 <\gamma_I(x) <1$ and $0 < \gamma_A(x) <1$ are the age-dependent recovery rates of infected and asymptomatic individuals, respectively. Consequently, symptomatic infected of age $x$ are removed by the epidemic dynamics through isolation (in hospital or at home) at a removal rate $\gamma_I(x)$ (with typical delay $\gamma_I(x)^{-1}$), while asymptomatic people are removed from the epidemic dynamics mostly through spontaneous recovery, at a recovery rate $\gamma_A(x) > \gamma_I(x)$, thus after a typical time $\gamma_A(x)^{-1} < \gamma_I(x)^{-1}$. As remarked in Ref.  \cite{Gaeta}, the possibility to detect asymptomatic individuals could leads to their isolation before healing, and thus to a reduction in  $\gamma_A(x)^{-1}$ .

As briefly explained in the Introduction, the nature of the COVID-19 epidemic is such that the functions $\gamma_I(x)$ and $\gamma_A(x)$ manifest a behavior \textcolor{black}{inversely} proportional to the age variable $x$. This expresses the fact that the TVC in both classes is increasing with age. Opposite behavior can be assumed for the function $\xi(x)$, since in this case the percentage of asymptomatic individuals has been noticed to decrease with age. In the next section, we will discuss once $\gamma_I(x)$ and $\gamma_A(x)$ are recovered from the data (with a certain uncertainty) and are inserted in our model, how the function $\xi(x)$, which is unknown, can be used to fit the model with the real observations.

We discuss now the terms $Q_J$, $J\in \{S,I,A,R\}$ responsible for the social interactions between individuals. We assume that the social contact variable thermalizes at a very rapid scale, i.e. that independently on the initial state the number of average daily contacts classified by age can be expressed by a given distribution, that is, as detailed next, a Gamma distribution,\cite{Sta} fixed with respect to time. More precisely only the mean of such distribution can change with respect to time as a result of the pandemic evolution, the shape remaining unchanged. In particular in the Appendix~\ref{sec:kinetic} we discuss the closure of the kinetic system \eqref{siar-gamma} around the local equilibrium of the social contact variable, which has been shown to be well-fitted by a Gamma function distribution \cite{Plos,DPeTZ}. By resorting to this closure, as shown later, we recover a continuum nonlinear system for the compartment's densities which describes the evolution in time of the populations of susceptible, asymptomatic, infected and recovered individuals in terms of their age. This system, however, retains memory of the effects of the underlying social contact variable in the spreading of the infection through the bilinear interaction operator quantifying the interactions between susceptible and infected individuals.

We assume then as in Ref.  \cite{DPeTZ} that the operators $Q_J$, $J\in \{S,I,A,R\}$ in \fer{siar-gamma} are Fokker--Planck type operators characterizing the relaxation towards an equilibrium state of the distribution of social contacts in the various classes in terms of repeated interactions \cite{DT,To3}. The mathematical justification, the details and the motivations leading to such results are detailed in the Appendix \ref{sec:kinetic}. These operators are expressed by the joined action of a diffusion, with a linearly variable coefficient and a linear drift
\be\label{FP-ope}
Q_J(f_J)(x,v,t)=  \frac{\partial}{\partial v}\left[ \nu \frac{\partial}{\partial v} (v f_J(x,v,t)) +  \left(\frac v{\bar m_J(x,t)} -1\right)f_J(x,v,t)\right],
\ee
complemented with \emph{no-flux} boundary conditions at the point $v=0$ 
\begin{equation}\label{bc}
\nu \frac{\partial}{\partial v} (v f_J(x,v,t)) +\left(\frac v{\bar m_J(x,t)} -1\right) f_J(x,v,t)=0\Big|_{v=0} = 0,
\end{equation}
for $J\in\{S,I,A,R\}$, and a suitable rapid decay of the densities $f_J(x,v,t)$ for $v\rightarrow +\infty$ \cite{FPTT16}.  In \fer{FP-ope}, the parameter $\nu$ is a positive constant characterizing the intensity of the diffusion operator, while $ \bar m_J(x,t)>0$ represents the observed \emph{mean number} of contacts at time $t\ge 0$ of the population of the $J$-th class of age $x$ at time $t \ge 0$ and can be extrapolated from data. Hence, the constant $\nu$ is a measure of the deviation of the population of a class from its observed average value. Note that, in absence of an external reason, the quantity $\bar m_J(x,t)$ is the same for the whole population, and it is expected to not depend on time, so that $ \bar m_J(x,t) = \bar m(x)$ \cite{Plos}.  On the contrary, in the presence of an epidemic, the observed mean numbers of daily contacts $\bar m_J(x,t)$ reasonably \textcolor{black}{tend} to decrease in time, even in absence of an external lockdown intervention, in reason of the perception of danger linked to social contacts that people manifest \cite{Bon}.

It is now important to notice that the operators $Q_J$ act only on the social contact variable $v$ and that they are \emph{mass and mean preserving}, that is such that 
\[
\dfrac{d}{dt}\int_{\R_+}\varphi(v)\,Q_J(f_J)(x,v, t)\,dv  = 0,
 \]
if $\varphi(v) =1, v$. Moreover, the kernel of the operators $Q_J$ contains the evoked Gamma probability densities
\be\label{gamma-J}
f_J^\infty(v;\bar c_J(x,t), \nu) =  \left(\frac \nu{\bar m_J(x,t)}\right)^\nu \frac 1{\Gamma\left(\nu \right)} v^{\nu -1}
\exp\left\{ -\frac\nu{\bar m_J(x,t)}\, v\right\},
 \ee 
of unitary local density. It is worth to mention that the Gamma densities \fer{gamma-J} are characterized by mean value and variance given by
\[
\begin{split}
 &\int_{\R^+} v\, f_J^\infty(v;\bar c_J(x,t), \nu) \,dv= \bar m_J(x,t)\\
& \int_{\R^+} (v-\bar m_J(x,t) )^2 \, f_J^\infty(x;\theta, \nu) \, dx =\nu\,\bar m_J(x,t)^2.
\end{split}\]
Given this equilibrium distribution it is important to highlight the meaning of the constant $\nu$, which characterizes the size of the variance of the functions in the kernel of the operators $Q_J$, $J\in \{S,I,A,R\}$. In fact, since the Gamma-type functions \fer{gamma-J} represent the equilibrium distribution of social contacts of the population, then, in presence of rigid lockdown measures, a small value of $\nu$ characterizes a population which is fully respecting  the measures themselves.

 \begin{remark} If the functions and $\kappa(v,w)$, $\gamma_I(x)$, $\gamma_A(x)$ and $\xi(x)$ are assumed constant, and $\CU(x) =\CI$, integration of system \fer{siar-gamma} with respect to the pair $(x,v)$ gives the system of differential equations for the mass fractions $J(t)$, $ J \in \{S,I,A,R\} $, considered by Gaeta in Ref.  \cite{Gaeta}, obtained by identifying the two classes of removed originating from infected and asymptomatic populations. Also we obtain a particular case of the system introduced in Ref.  \cite{LTB}, which assigns different rates of transmission to infected and asymptomatic individuals.
\end{remark}
 
 \begin{remark} The solution of the kinetic system \fer{siar-gamma} allows to obtain the evolution in time of the distribution functions  $f_J(x,v,t)$, $J\in \{S,I,A,R\}$, which depend on both the age variable $x$ and the social contact variable $v$. However, the variation of these distributions is heavily dependent on the action of the operators $Q_J$, $J\in \{S,I,A,R\}$, and this action is inversely proportional to the relaxation constant $\tau$. The constant $\tau$ plays an important role in the time evolution of the model \fer{siar-gamma}, since it relates the time scale of the epidemic evolution with that of the statistical  formation of social contacts. Small values of the constant $\tau$ will correspond to a fast adaptation of people to a steady situation. Hence, since it is highly reasonable to assume that  adaptation of people is faster \textcolor{black}{than the epidemic dynamics},  following Ref.  \cite{DPeTZ} we will assume $\tau \ll 1$. The consequences of this choice  can be fully understood by resorting once again to the analogies with classical kinetic theory of rarefied gases (see Ref.  \cite{PT13} and Appendix \ref{sec:kinetic}).
  \end{remark}
 \subsection{The macroscopic S-SIAR model}\label{sec:splitting}
Resorting to the discussion of the previous section, to Appendix \ref{sec:kinetic} for a mathematical justification and supposing $\tau\ll 1$, we integrate both sides of equations in \fer{siar-gamma} with respect to $v$. Since the Fokker-Planck type operators \fer{FP-ope} are mass-preserving, we obtain the system for the evolution of the local densities $\rho_J(x,t)$, defined in \fer{densi}, $J\in\{S,I,A,R\}$ 
\begin{equations}\label{siar-inte}
\frac{\partial }{\partial t} \rho_S(x,t)=& -  \beta \, c_S(x,t)\int_{\CU(x)} \left(c_I(y,t) + c_A(y,t)\right)\, dy - \bar\chi(x,t)
\\
\frac{\partial }{\partial t} \rho_I(x,t) = &\beta\, \xi(x)  \, c_S(x,t)\int_{\CU(x)} \left(c_I(y,t) + c_A(y,t)\right)\, dy  - \gamma_I(x) \rho_I(x,t) 
\\
\frac{\partial }{\partial t} \rho_A(x,t) =& \beta\,(1- \xi(x))  \, c_S(x,t)\int_{\CU(x)} \left(c_I(y,t) + c_A(y,t)\right)\, dy  \\
&- \gamma_A(x) \rho_A(x,t)(x,v,t) 
\\
\frac{\partial }{\partial t} \rho_R(x,t) =& \gamma_I(x)  \rho_I(x,t) + \gamma_A(x)  \rho_A(x,t) +  \bar\chi(x,t)
 \end{equations}
with $\bar\chi(x,t)=\int_{\R^+}\chi(x,v,t)dv$.

Clearly, system \fer{siar-inte} is not closed, since the evolution of the local densities $\rho_J(x,t)$, defined in \fer{densi}, \textcolor{black}{ $J\in\{S,I,A,R\}$}, depends on the mean numbers $c_J(x,t)$. Then, the closure of system \fer{siar-inte} can be obtained by resorting to the Gamma equilibria \fer{gamma-JJ}. In fact, as outlined, the typical time scale involved in the social contact dynamics is $\tau\ll1$ which identifies a faster adaptation of individuals to social contacts with respect to the evolution time of the epidemic disease. For example, the choice of the value $\tau \ll1$ pushes the distribution function $f_S(x,v,t)$ in the first equation of system \fer{siar-inte} towards the Gamma equilibrium density with a local density  $\rho_S(x,t)$ and local mean density  $\bar m_S(x,t)$ and the same for the other compartments. 
The next step needed to have a complete closed system is to quantify in a correct way the mean value $\bar m_S(x,t)$, as for instance explained in Ref.  \cite{DPeTZ}. This follows by assuming that, that due to the presence of the epidemic, the population tends to reduce the typical average number of contacts which exhibits in standard situations. This can happen due to two main reasons: on voluntary basis for preventing contagion or by authorities decision through lockdown measures. This average reduction effect can be introduced by assuming that the local mean densities of daily contacts of susceptibles and asymptomatic individuals $\bar m_J(x,t)$,   \textcolor{black}{ $J\in\{S,I,A,R\}$}, of the Gamma-type equilibria \fer{gamma-JJ} depends on both the proportion of infected and the age of individuals through
 \be\label{cons}
 \bar m_J(x,t) =  m(x)\,H_J(x, I(t)) ,
 \ee
where $m(x)$ is the constant in time local mean density in an epidemic free situation (such the one discussed in Ref.  \cite{Plos}). The functions $H_J(\cdot, r)$ are non increasing with respect to $r$, $0\le r \le 1$, starting from $H_J(\cdot, 0) = 1$ and they characterize the decreasing of daily contacts in relation to the evolution of the pandemic. Note that the functions $H_J$ \textcolor{black}{do not depend} on the number of asymptomatic individuals, in reason of the fact that the size of this class is difficult to quantify, and the main indicator of the spreading of the epidemic is usually given by the amount in time of registered infectious individuals. Moreover, the mean number of social contacts for the class of recovered is assumed constant in time. This is due to the fact that the behavior of epidemic does not depend on it. With the above hypothesis, we finally obtain the closed system
 \begin{equations}\label{sir-closed}
& \frac{\partial}{\partial t}   \rho_S(x,t)= - \Lambda(x,t) - \bar\chi(x,t)\\
& \frac{\partial }{\partial t}  \rho_I(x,t)= \xi(x)  \, \Lambda(x,t) - \gamma _I(x)\rho_I(x,t), 
 \\
 & \frac{\partial }{\partial t} \rho_A(x,t) = (1- \xi(x))   \, \Lambda(x,t) - \gamma _A(x)\rho_A(x,t), 
 \\
& \frac{\partial }{\partial t} \rho_R(x,t) =  \gamma_I(x)  \rho_I(x,t) + \gamma_A(x)  \rho_A(x,t) + \bar\chi(x,t).
\end{equations}
In \fer{sir-closed} the function $\Lambda(x,t)$ is the age-dependent incidence rate function at time $t >0$, given by
 \be\label{age-inci}
 \begin{split}
 \Lambda(x,t)  =& \beta\, m^2(x)  \,H_S(x,I(t)) \, \rho_S(x,t) \\
 &\times \int_{\CU(x)}(H_I(y,I(t))\rho_I(y,t)+H_A(y,I(t))\rho_A(y,t))\, dy.
 \end{split}
\ee
We refer to this model to as the S-SIAR model, where SIAR in the standard acronym for a compartment model with asymptomatic and S states for the fact that our model takes into account ages and social contacts \textcolor{black}{of} individuals. 
 
Finally, in agreement with the analysis of Ref.  \cite{DPeTZ}, for a given constant $N \gg 1$ we assume that the functions $H_J(x,r)$ are given in Ref.  \cite{KM}
 \be\label{ese}
 \begin{split}
 H_J(r) &= \frac{\mu_J(r) }{\sqrt{1+ N r} }, \quad 0\le r \le 1, \\
\mu_I(r) &= \kappa \mu_S(r), \qquad \kappa \in [0,1]\\
\mu_A(r) &=\mu_S(r). 
 \end{split}
 \ee
 which describe a possible way in which, in presence of the spread of the disease, the susceptible and asymptomatic populations tend to reduce in the same way the mean number of daily social contacts whereas the contacts of the population of infected is reduced by a further quantify  $\kappa\in [0,1] $. All the above introduced functions depend on the age, where this dependence is expressed by the function $0 <\mu_J(x)<1$. In addition to the form \eqref{ese}, we can also consider the function
\be\label{ese1}
H_J(t,r) = \frac{\mu_J(x) } {\sqrt{1+ Nr(t) \displaystyle\int_0^t r(s)ds} }, \quad 0\le r \le 1.
\ee
This function satisfies the same properties \textcolor{black}{of} \eqref{ese} in terms of the incidence rate requests detailed in Ref.  \cite{KM} and takes into account memory effects on the population's behavior. In fact, people may adapt their life style in terms of possible daily contacts to answer to the actual pandemic situation as expressed by \eqref{ese}, however further effects based on the global dynamics of the disease can influence the choices of the population.

\subsection{Some preliminary numerical experiments}\label{sec:numerics}
In this section, before using the dataset to determine the values of the parameters intervening in the model and provide reliable results about the pandemic, we discuss several academic numerical examples with the scope of providing extensive evidences of the capabilities of the S-SIAR model (\ref{sir-closed}) in describing different epidemic situations. In details, we provide results starting from different hypothesis on the contact function $H$ and for a given set of initial data observing the different scenarios provided by the model. 

\begin{figure}\centering
    {
        \includegraphics[scale = 0.4]{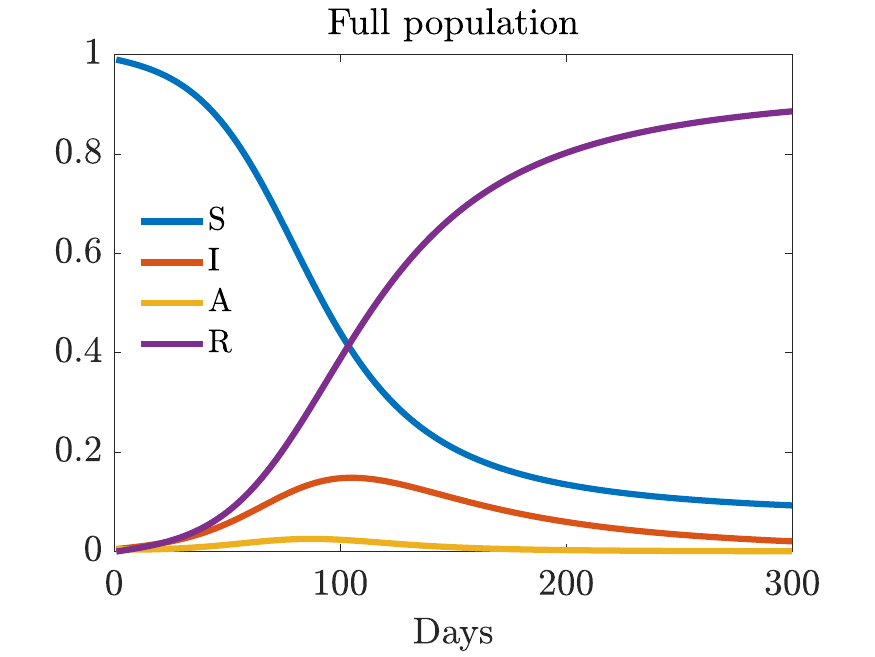} 
        \includegraphics[scale = 0.4]{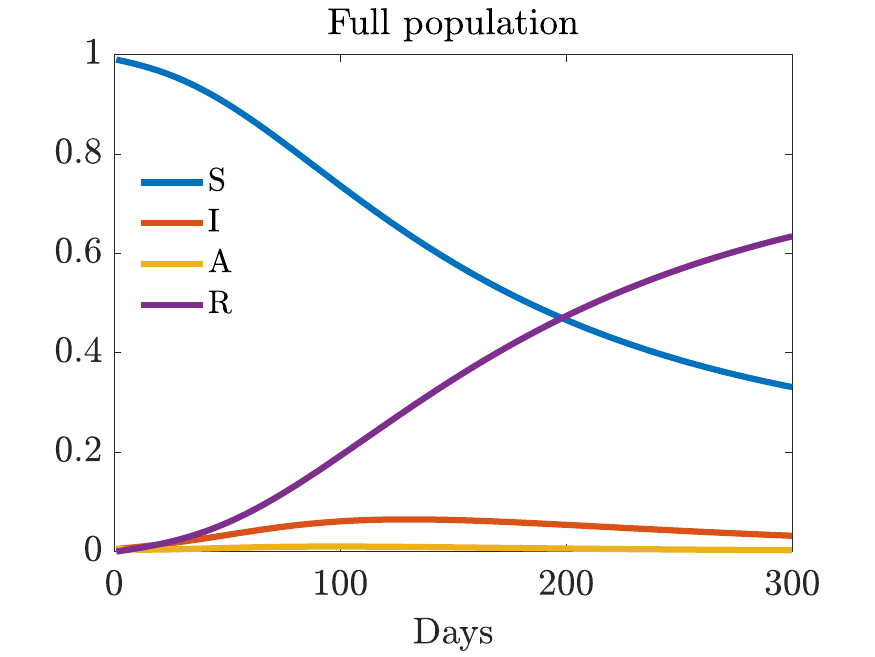} \\
        \includegraphics[scale = 0.4]{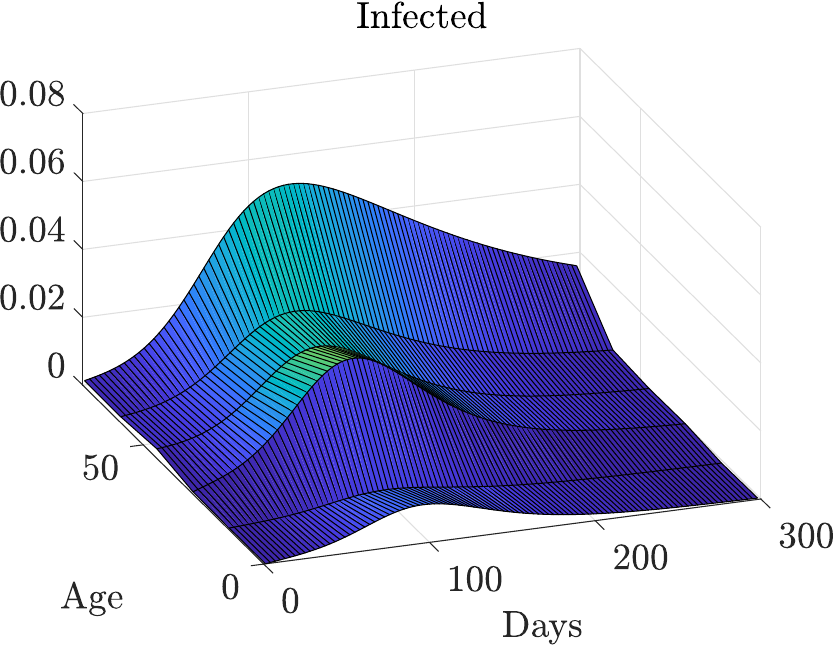} 
        \includegraphics[scale = 0.4]{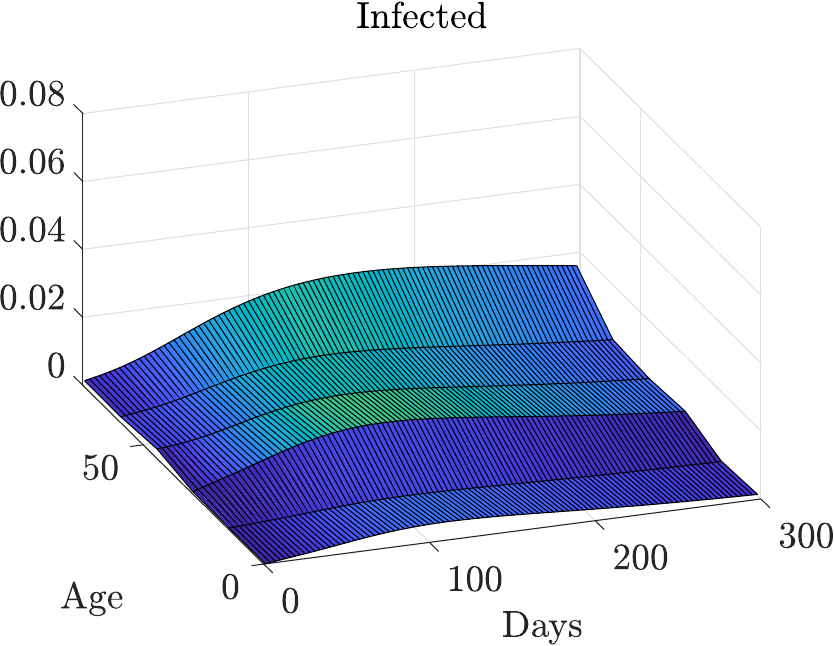}\\
        \includegraphics[scale = 0.4]{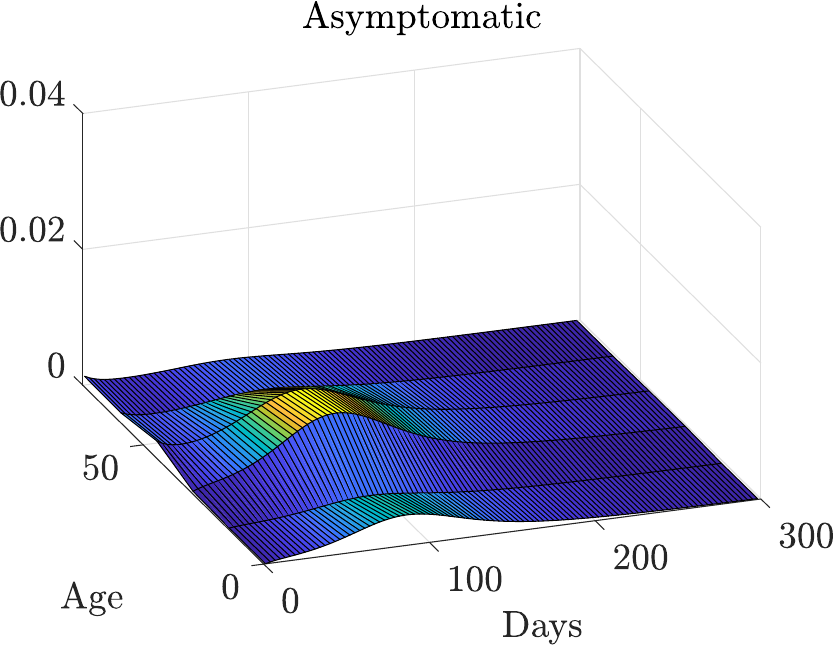} 
        \includegraphics[scale = 0.4]{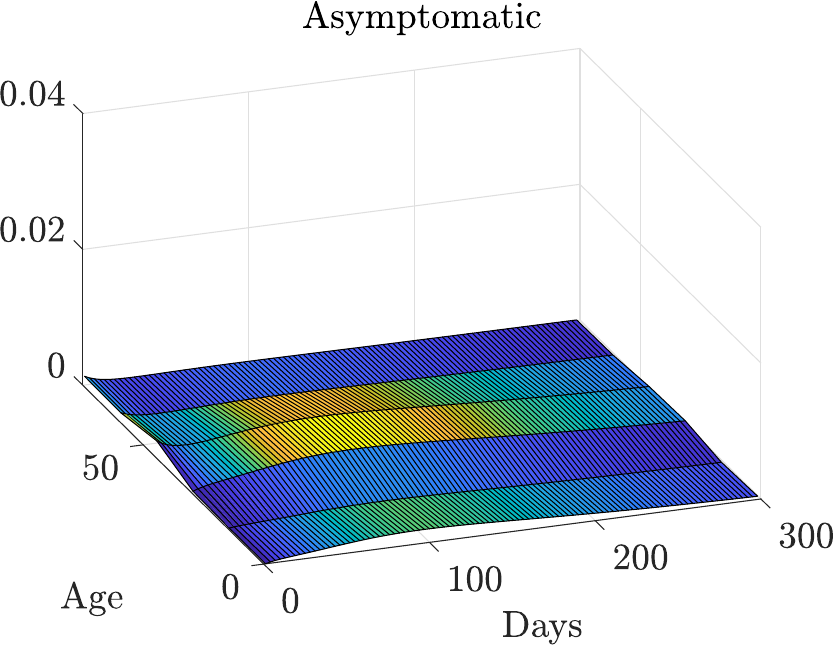}\\ 
    }
    \caption{Test 1A. Time evolution of the S-SIAR model. Left images report the case in which the contact function is fixed to one. Right images show the case $H(x,t)$ given by (\ref{ese}).From top to bottom, full population dynamic, number of  Infected and Asymptomatic as a function of age and time.}\label{fig:test1_1}
\end{figure}

\begin{figure}\centering
    {
        \includegraphics[scale = 0.4]{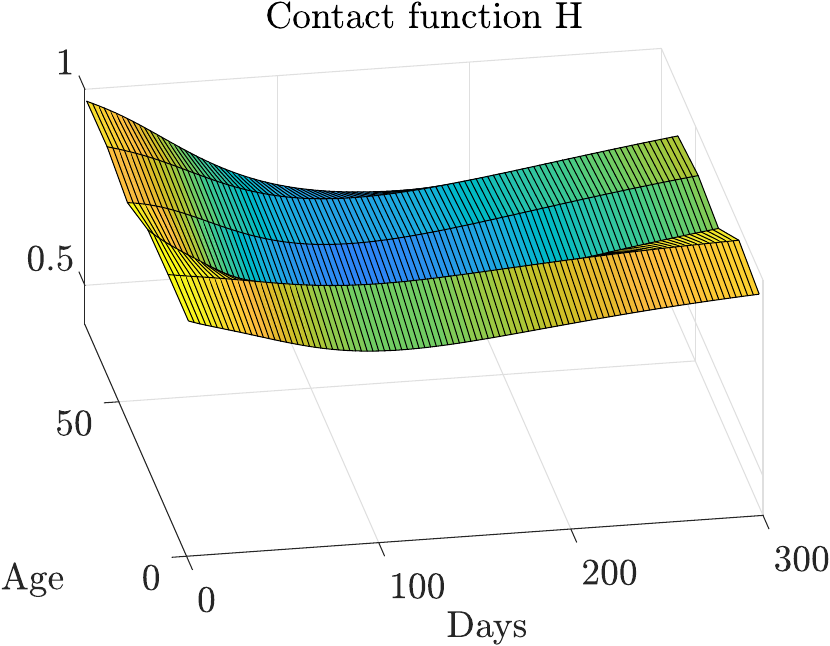} 
        \includegraphics[scale = 0.4]{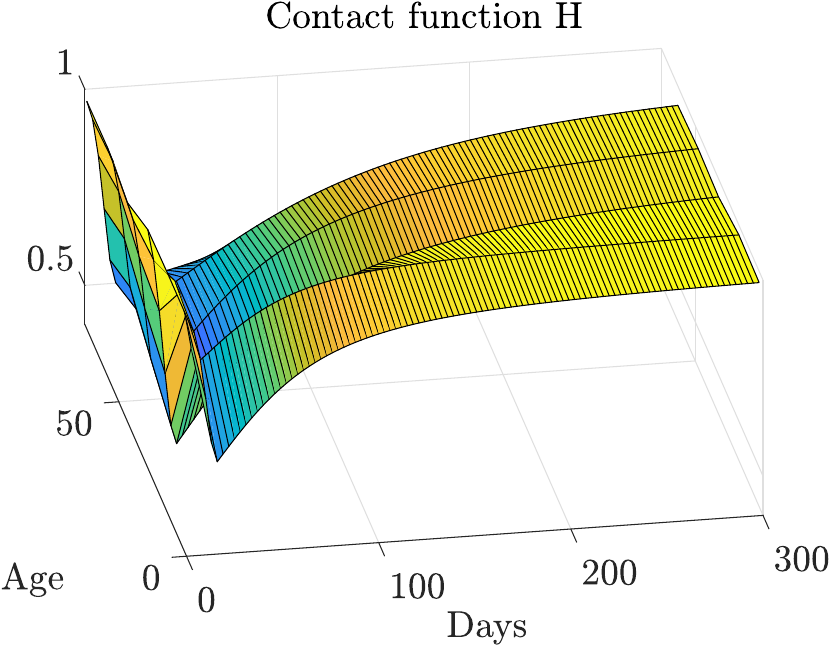}
        \includegraphics[scale = 0.4]{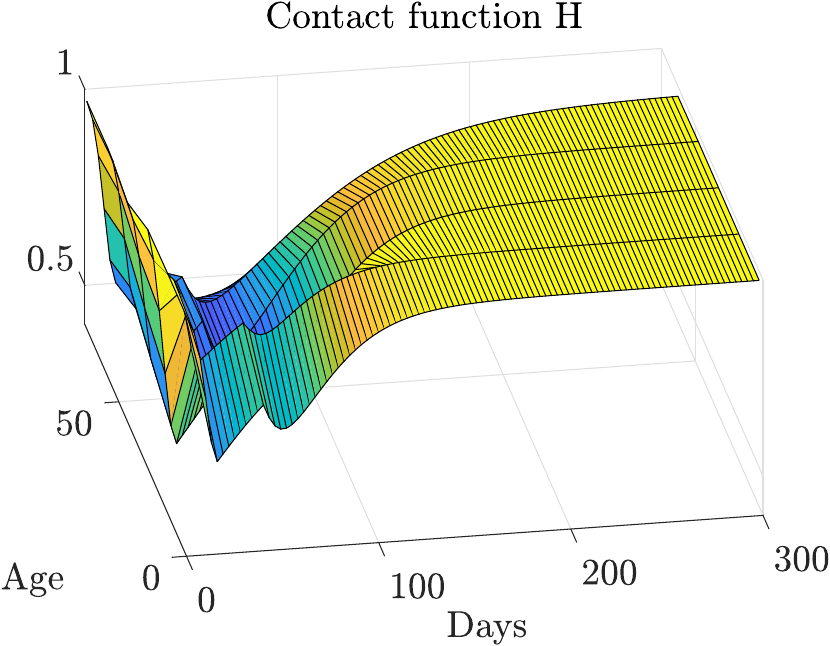}
    }
    \caption{Test 1C. Time evolution of the social contact function $H(x,t)$. Top left shows the case of Test 1 of Figure \ref{fig:test1_1} right. Top right and bottom show the case of Test 2 for the two different lockdown strategies: lockdown up to the end of the epidemic or time limited lockdown.}\label{fig:test1_3}
\end{figure}

We start by fixing an initial value of susceptible, infected, asymptomatic and recovered individuals for six different classes of age. \textcolor{black}{For these initial tests we do not consider vaccinations}, i.e. we fix $\bar\chi(x,t)=0$. The chosen classes are the following:
\be 
\begin{split}
&0<x\le 18, \quad 19\le x\le 24, \quad 25\le x\le 49,\\
 &\quad 50\le x\le 64, \quad 65\le x\le 74, \quad x\ge 75.
\end{split}\ee 
These are the same classes provided by the dataset at our disposal. These also well represent the standard classification of the average severity at which the COVID-19 acts over the population. The initial fractions of people belonging to a given class is 
\begin{equation*}
    \left\{ \begin{array}{ll}
        N_1=0.14 & 0<x\le 18\\ 
        N_2=0.065 & 19\le x\le 24 \\
        N_3=0.315 & 25\le x\le 49\\ 
        N_4=0.23 & 50\le x\le 64 \\
        N_5=0.12 & 65\le x\le 74\\ 
        N_6=0.13 & x\ge 75 \\
    \end{array} \right. 
\end{equation*}
summing up to one. This choice reflects the distribution of ages in the Pavia province, Italy, in 2020. At the same time the initial number of Infected and Asymptomatic is given by 
\begin{equation*}
    \left\{ \begin{array}{ll}
        I_1(t=0)=A_1(t=0)=10^{-4}N_2 & 0<x\le18\\ 
        I_2(t=0)=A_2(t=0)=10^{-4}N_3 & 19\le x\le 24 \\
        I_3(t=0)=A_3(t=0)=5\ 10^{-4}N_1 & 25\le x\le 49\\ 
        I_4(t=0)=A_4(t=0)=10^{-2}N_4 & 50\le x\le 64 \\
        I_5(t=0)=A_5(t=0)=10^{-2}N_5 & 65\le x\le 74\\ 
        I_6(t=0)=A_6(t=0)=10^{-2}N_6 & x \ge 75 \\
    \end{array} \right. 
\end{equation*}
Correspondingly, the initial number of Recovered is fixed to zero for all classes. The epidemiological parameters are finally $m_{\beta_i}=\beta m(x_i)=0.1, \ i=1,..,6$ and
\begin{equation*}
    \left\{ \begin{array}{llll}
        \gamma_{I,1}=0.05 & \gamma_{A,1}=0.14 & \xi_{I,1}=0.5 & 0<x\le 18\\ 
        \gamma_{I,2}=0.05 & \gamma_{A,2}=0.14 & \xi_{I,2}=0.5 & 19\le x\le 24 \\
        \gamma_{I,3}=0.045 & \gamma_{A,3}=0.126 & \xi_{I,3}=0.6 & 25<x\le 49\\ 
        \gamma_{I,4}=0.04 & \gamma_{A,4}=0.112 & \xi_{I,4}=0.6 & 50<x\le 64 \\
        \gamma_{I,5}=0.025 & \gamma_{A,5}=0.112 & \xi_{I,5}=0.8 & 65<x\le 74\\ 
        \gamma_{I,6}=0.01 & \gamma_{A,6}=0.112 & \xi_{I,6}=0.9 & x\ge 75 \\
    \end{array} \right. 
\end{equation*}
Given these values, we observe the different results obtained by the S-SIAR model (\ref{sir-closed}) in Figure \ref{fig:test1_1}. This figure shows on the top the time evolution of the epidemic in terms of the full population for two distinct cases: the first, on the left, uses a contact function $H(x,t)=1$, while the second, on the right, uses $H(x,t)=1/\sqrt{1+50 I(x,t)}$. The central and the bottom images shows the details of the Asymptomatic and Infected per class. From this figure, it is clear that taking into account a contact function which naturally decreases as the number of Infected increases permits to diminish the impact of the pandemic. The Figure \ref{fig:test1_3}, top left, shows the time evolution of the contact function $H(x,t)$ per age. 

\begin{figure}\centering
    {
        \includegraphics[scale = 0.4]{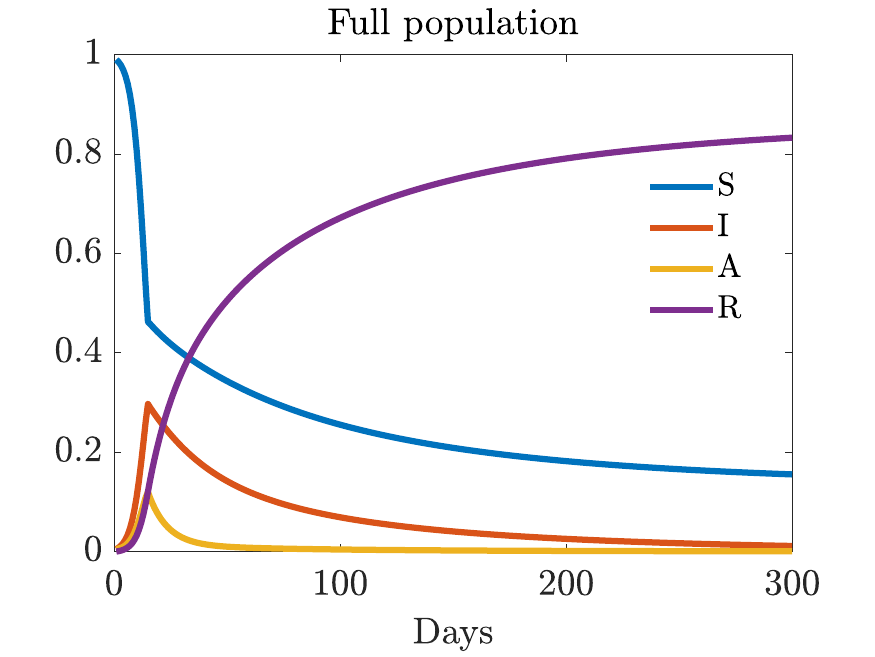}
        \includegraphics[scale = 0.4]{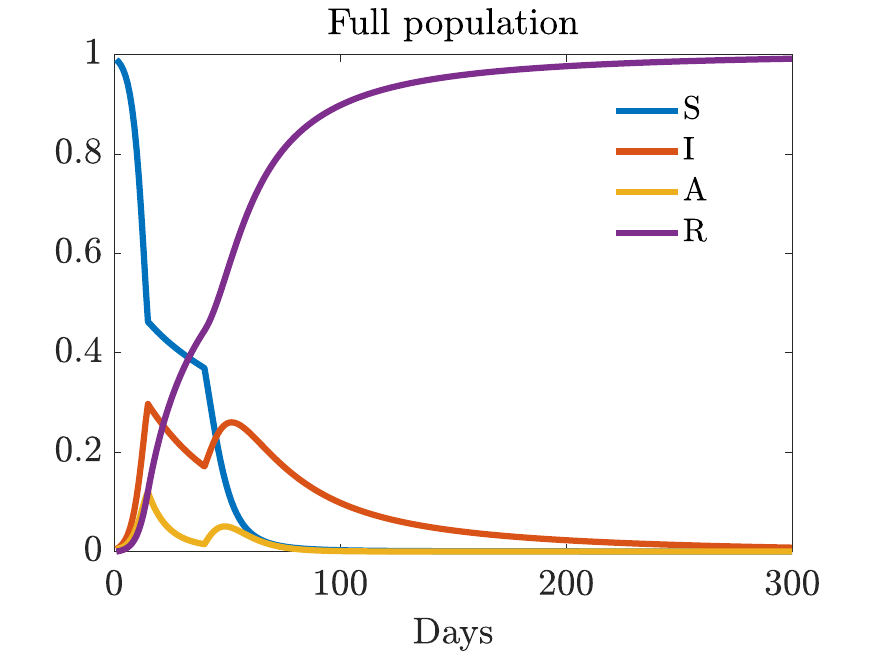} \\ 
        \includegraphics[scale = 0.4]{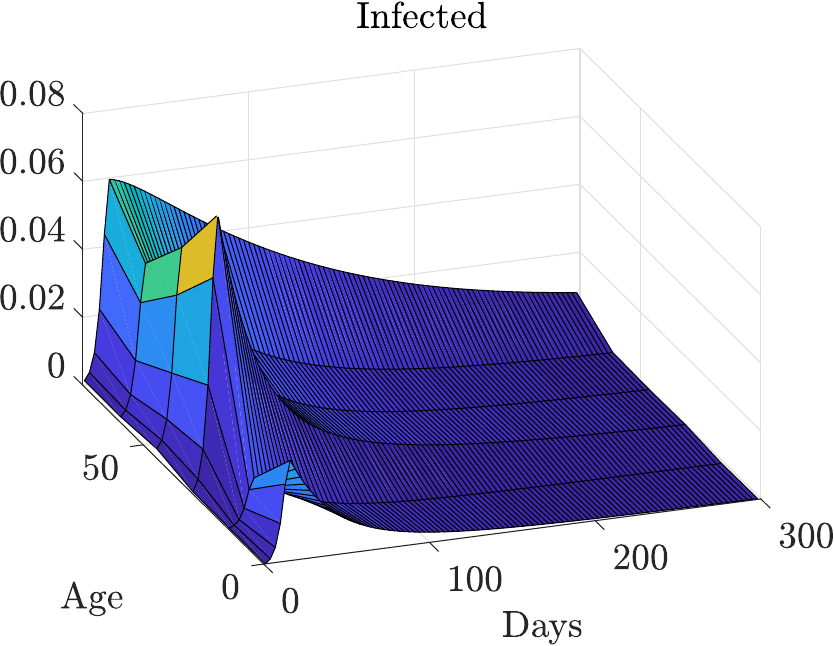} 
        \includegraphics[scale = 0.4]{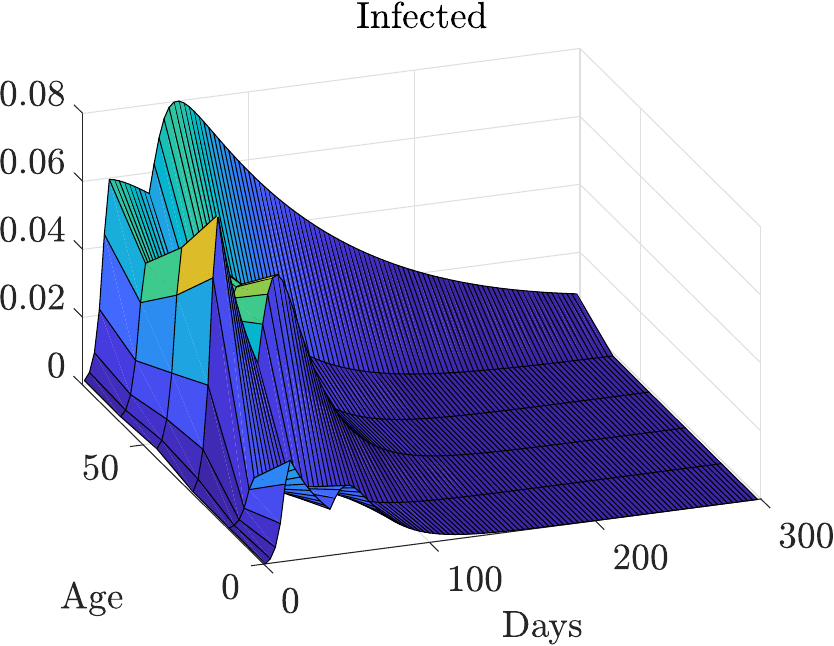}\\      
        \includegraphics[scale = 0.4]{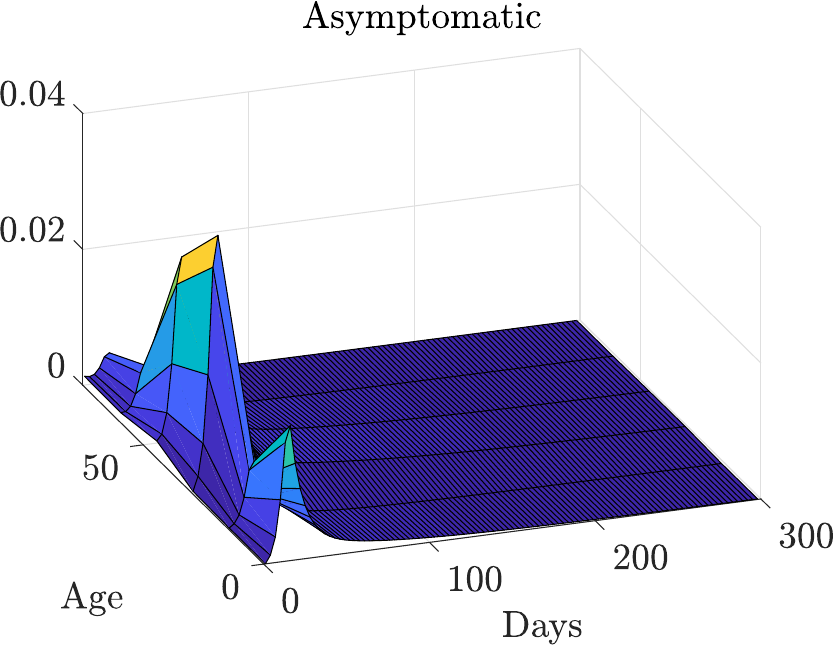} 
        \includegraphics[scale = 0.4]{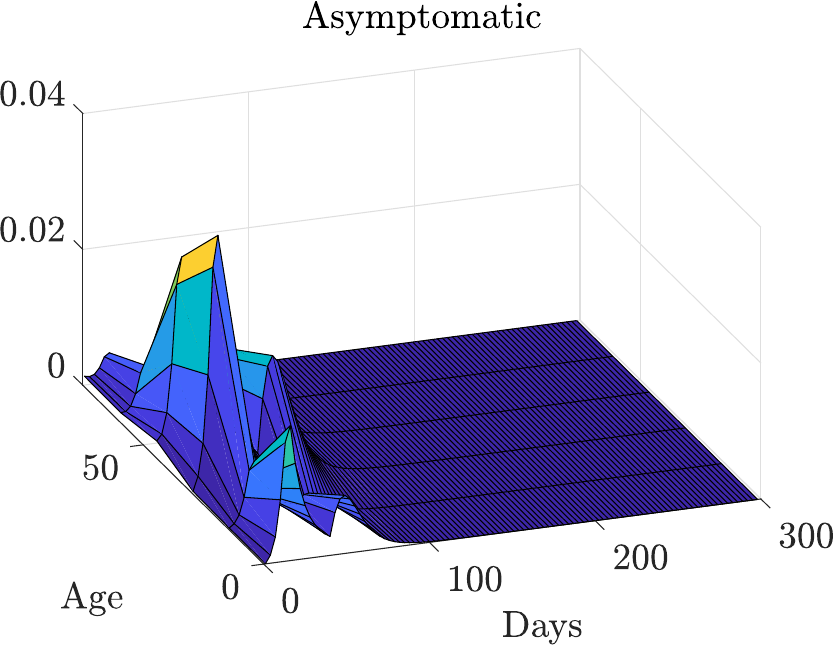}\\
    }
    \caption{Test 1B. Time evolution of the S-SIAR model. Left images report the case in which a lockdown is imposed up to the end of the epidemic. Right images show the effect of a lockdown imposed for a time window of 25 days. From top to bottom, full population dynamic, number of  Infected and Asymptomatic as a function of age and time.}\label{fig:test1_2}
\end{figure}

In Figure \ref{fig:test1_2}, the results for other two different scenarios are reported. For these cases, the contacts rates have been elevated to $m_{\beta_i}=0.4$, $ i= 1,\dots,6$. The first scenario, shown on the left images, assumes a change in the behavior of the population caused by some hypothetical restrictions on the mobility imposed by the Authorities starting 15 days after the occurrence of pandemic. The second scenario, assumes analogous restrictions on mobility but only over a time window of 25 days, starting 15 days after the onset of the pandemic and stopping after 40 days. In this second case, we clearly see a second wave of the epidemic raising again. Finally, Figure \ref{fig:test1_3} top left and bottom, illustrates the time evolution of the social contact function for these two cases divided per age. The results presented show that the S-SIAR model is capable of taking into account the role played by the number of average daily contacts with respect to the spread of the disease and in a different manner depending on the age of individuals. With these results in mind, we will focus in the next section on realistic scenarios matching the data with the model.

\section{Calibration and sensitivity of the S-SIAR model}
\label{sec:calibration}
In this section, we propose a strategy for the estimation of relevant epidemiological parameters of the introduced age-dependent S-SIAR model which takes into account the described dataset provided by the Health Protection Agency of the Province of Pavia, Italy. In the second part, we show some forecasting and vaccination scenarios. The calibration of epidemiological models represent in general a difficult task and require special care both at the numerical level and in data assimilation. Unfortunately, available data related to the spreading of a disease represent an incomplete picture of the phenomenon under consideration. Indeed, according to recent results of the serological campaigns promoted in Italy for the early phase of the pandemic, around $80\%$ of infected have not been registered in official datasets.\footnote{\texttt{Istituto Nazionale di Statistica \\\texttt{https://www.istat.it/it/files//2020/08/ReportPrimiRisultatiIndagineSiero.pdf}}}
The necessity to face highly heterogeneous data, together with their unavoidable incompleteness, makes the estimation of relevant epidemiological parameters challenging. For this reason, we tackle the problem of parameter estimation using a strategy based on uncertainty quantification, where population heterogeneity for each available age segment is considered in a statistical setting. For related approaches we mention the recent works  \cite{APZ,Cetal,Chowell}. This permits to furnish results in terms of expected mean values and confidence bands.
It is worth to mention that to handle more efficiently the uncertainties in the dynamics due to heterogeneity of recoveries we have adopted a stochastic collocation approach based on stochastic Galerkin methods  (see Ref.  \cite{X} for an introduction and Ref.  \cite{APZ,Rob,Zanella} for examples related to the spread of a disease).

We summarize the parameters that have to be determined in \eqref{sir-closed} in order to have a model which permits to give reliable results. Coherently with the data at our disposal, we suppose the vaccination campaign not started yet. The parameters are the incidence rates $\Lambda(x,t)$, the recovery rates $\gamma_A(x),\gamma_I(x)$, the fraction of asymptomatic $\xi(x)$ per class of age and the initial pandemic situation. From the dataset we are able to obtain, as explained in Section \ref{sec:stat}, the distribution of the recovery rates $\gamma_A(x),\gamma_I(x)$  and the initial pandemic situation. The incidence rates and the fraction of asymptomatic are instead information that, as detailed next, are extrapolated from the fitting with the dataset.

\subsection{Extrapolation of the incidence rates and fraction of asymptomatic}
 
In the following we present a bilevel approach which is based on two main optimization horizons that have been chosen in relation to the adopted regional measures for the second epidemic wave. In details, on October 8th 2020 a national act established that the use of masks were mandatory for any indoor and outdoor activities, hence several other specific measures were implemented in the following week and curfew has been imposed in the Lombardy region starting from October 22nd 2020 for which any not justified travel was forbidden during evening hours. This curfew has been extended later at a national level together with other more restrictive measures starting from November 3rd 2020 also regarding the closure of schools and universities. 

We consider then first the time interval $[t_0,t_\ell]$, where $t_0$ is October 8th 2020 and $t_\ell$ October 22nd, in which more restrictive measures started in Lombardy. Hence, the distribution of the recovery rates $\gamma_{I,A}(x)>0$ based on the analysis on the TVC reported in Section \ref{sec:stat} are known. In particular in Table \ref{tab:age_param}, we reported the observed TVC for each age segment of the dataset which establish a clear age-dependent pattern. However, as data show, the recovery rates are not fixed but instead distributed accordingly to a beta law. For this reason, we construct a 2D parameter $\gamma(\z,x)$ where $\z = (z_1,z_2)$ and $z_1,z_2$ are independent random variables with different estimated Beta distributions
\[
z_1 \sim B(\alpha_1,\beta_1),\qquad z_2\sim B(\alpha_2,\beta_2), 
\]
with $\alpha_1 = 2.1$, $\beta_1 = 5.1$ and $\alpha_2 = 1.8$, $\beta_2 = 3.9$. Furthermore, in order to reproduce the estimated TVC we consider 
\[
\dfrac{1}{\gamma_I(x,z_1,z_2)}= 
\begin{cases}
h_{1,1} + z_1 h_{1,2}, \qquad x \le 49 \\
h_{2,1} + z_2 h_{2,2}, \qquad x>50, 
\end{cases}
\]
with $h_{1,1} = 5$, $h_{1,2} = 32$ and $h_{2,1} = 5$, $h_{2,2} = 40$ (see Table \ref{tab:age_param}). Coherently with Ref.  \cite{Gatto} we consider that the ratio between the recovery rates for asymptomatic and symptomatic is given by $\gamma_A = 2\gamma_I$. As a consequence of the introduced 2D uncertainty system \eqref{sir-closed} in absence of vaccines becomes
 \begin{equations}\label{sir-closed_z}
& \frac{\partial }{\partial t}\rho_S(\z,x,t) = - \Lambda(\z,x,t)  \\
& \frac{\partial }{\partial t} \rho_I(\z,x,t)= \xi(\z,x)  \, \Lambda(\z,x,t) - \gamma _I(\z,x)\rho_I(\z,x,t), 
 \\
 & \frac{\partial }{\partial t}\rho_A(\z,x,t) = (1- \xi(\z,x))   \, \Lambda(\z,x,t) - \gamma _A(\z,x)\rho_A(\z,x,t), 
 \\
& \frac{\partial }{\partial t} \rho_R(\z,x,t)=  \gamma_I(\z,x)  \rho_I(\z,x,t) + \gamma_A(\z,x)  \rho_A(\z,x,t),
\end{equations}
with 
\begin{equation}
    \label{age-inci_z}
    \begin{split}
    \Lambda(\z,x,t) =& m_\beta(\z,x)\,H_S(\z,x,t) \, \rho_S(\z,x,t)\\
    &\times \int_{\CU(x)}(H_I(\z,y,t)\rho_I+H_A(\z,y,t)\rho_A(\z,y,t))\, dy,
    \end{split}
\end{equation}
where $m_\beta(\z,x)$ contains the combined effects of the parameter $\beta$ in \ref{sir-closed} and of the average number of contacts $m(x)$ in absence of contagious and $H_I(\z,y,t) = \kappa H_S(\z,x,t)$.

We construct now a 2D sample $\{\gamma_{i,j}(x)\}_{i,j = 1}^M$ from $\gamma(x,z_1,z_2)$. The samples are obtained in a collocation setting through Gauss-Jacobi polynomials with $M = 5$ nodes. Therefore, for each $i,j = 1,\dots,M$ we consider the following optimization procedure. 
To determine $m_\beta(x) = m_{\beta_{i,j}}(x) \ge 0$ and the initial number of asymptomatic $ \rho_A(x,0) = {\rho_{A}}_{i,j}(x,0) \ge 0$, $t \in [t_0,t_\ell]$, we solve a least square problem based on the minimization of a cost functional $\mathcal J$ which takes into account the sum of two $L^2$ norms over the time horizon $[t_0,t_\ell]$. In particular, for each $i,j = 1,\dots,M$ we consider a cost taking into account the age-depedent reported number of infected $\hat \rho_I(x,t)$ and the reported recovered patients $\hat \rho_R(x,t)$, and the evolution of $\rho_I(x,t)$ and $\rho_R(x,t)$ given by the S-SIAR model \eqref{sir-closed} assuming $H_S = H_I \equiv 1$ and $\gamma(x) = \gamma_{i,j}(x)$. We are therefore interested in the solution of the constrained optimization problem 
\begin{equation}
\label{eq:min1}
\min_{m_\beta(x) \in [0,1], \rho_A(x,0)> 0} \int_{\mathcal I} \mathcal J(\hat \rho_I,\hat \rho_R,\rho_I,\rho_R) P(x)dx, \qquad  t \in [t_0,t_\ell], 
\end{equation}
subject to the dynamics given by \eqref{sir-closed} and where $P(x)$ is the age distribution of the province of Pavia. In \eqref{eq:min1} the cost functional $\mathcal J$ is a weighted convex combination between two relative $L^2([t_0,t_\ell])$ norms
\begin{equation}
\label{eq:J}
\mathcal J(\hat \rho_I,\hat \rho_R,\rho_I,\rho_R) = \mu \dfrac{\| \hat \rho_I(x,t) - \rho_I(x,t) \|_{L^2}}{\|\rho_I(x,t) \|_{L^2}} + (1-\mu)\dfrac{\| \hat \rho_R(x,t) - \rho_R(x,t) \|_{L^2}}{\|\rho_R(x,t) \|_{L^2}}, 
\end{equation}
with $\mu \in [0,1]$.


Once the mentioned relevant epidemiological parameters have been estimated before the introduction of the curfew in Lombardy, we can proceed with the second estimation of the function ${H_S}_{i,j}(x,t)= H_S(x,t)$ giving the limitation of contacts. This will furnish the complete evolution of the incidence rates $\Lambda(x,t)$ with time and age necessary to fit the model with the data. In addition, in this second optimization procedure, the parameters $\xi_{i,j}(x) = \xi(x) \ge 0$ in the time span $[t_\ell+1,T]$ are estimated which give the number of asymptomatic per age. As in \eqref{ese}, we consider the case where infected are recognized and quarantined. Therefore, the spreading of the epidemics, is supposed to be mainly due to the asymptomatic population. Anyway, as explained in Ref.  \cite{Zhang}, the infected population is likely to maintain a certain background contact rate in consequence of a portion of family contacts which cannot be suppressed. For this reason we fixed $\kappa = 0.65$ in our model meaning that the infected are still able to infect a certain number of individuals as experimentally observed. Finally, in order to reduce the fluctuations due to possible delays in the registration procedure which will appear in a day by day optimization strategy, we estimate the evolution of the function $H_S(x)$ and of the parameter $\xi(x)$ for a sequence of time steps (days) $t_i \in [t_i-\kappa_\ell, t_i+\kappa_r]$, $\kappa_\ell,\kappa_r\ge1$ chosen to cover a week of data, i.e. we average the fitting over a week. Thus, the following optimization problem has then finally been considered
\[
\min_{H_S(x) \in [0,1],\xi(x)} \int_{\mathcal I}\mathcal J(\hat \rho_I,\hat \rho_R,\rho_I,\rho_R)P(x)dx, \qquad t \in [t_i-\kappa_\ell,t_i+\kappa_r], 
\]
subject to \eqref{sir-closed}. The cost functional $\mathcal J$ has been defined in \eqref{eq:J} where the $L^2$ norms are computed in the time interval $[t_i-\kappa_\ell,t_i+\kappa_r]$. 

\begin{figure}
\centering
\includegraphics[scale = 0.33]{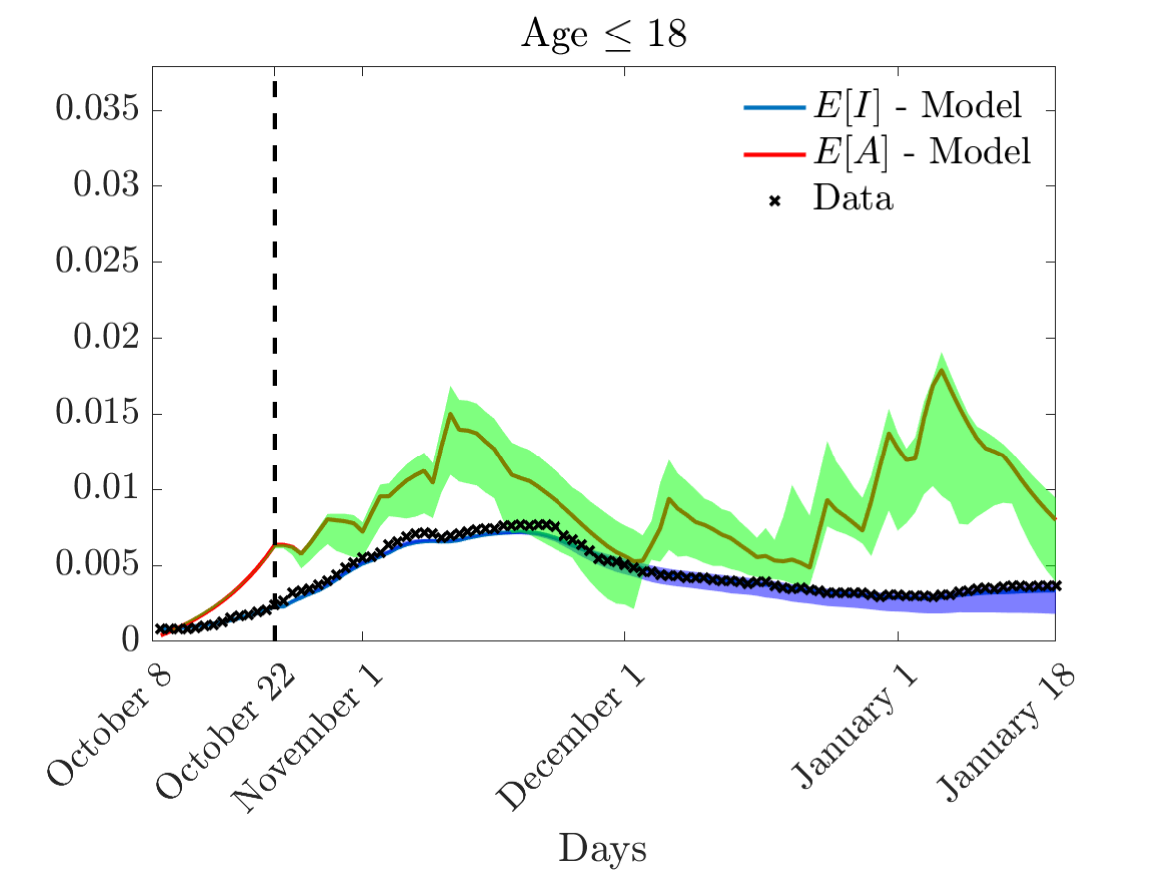}\hspace{-0.5cm}
\includegraphics[scale = 0.33]{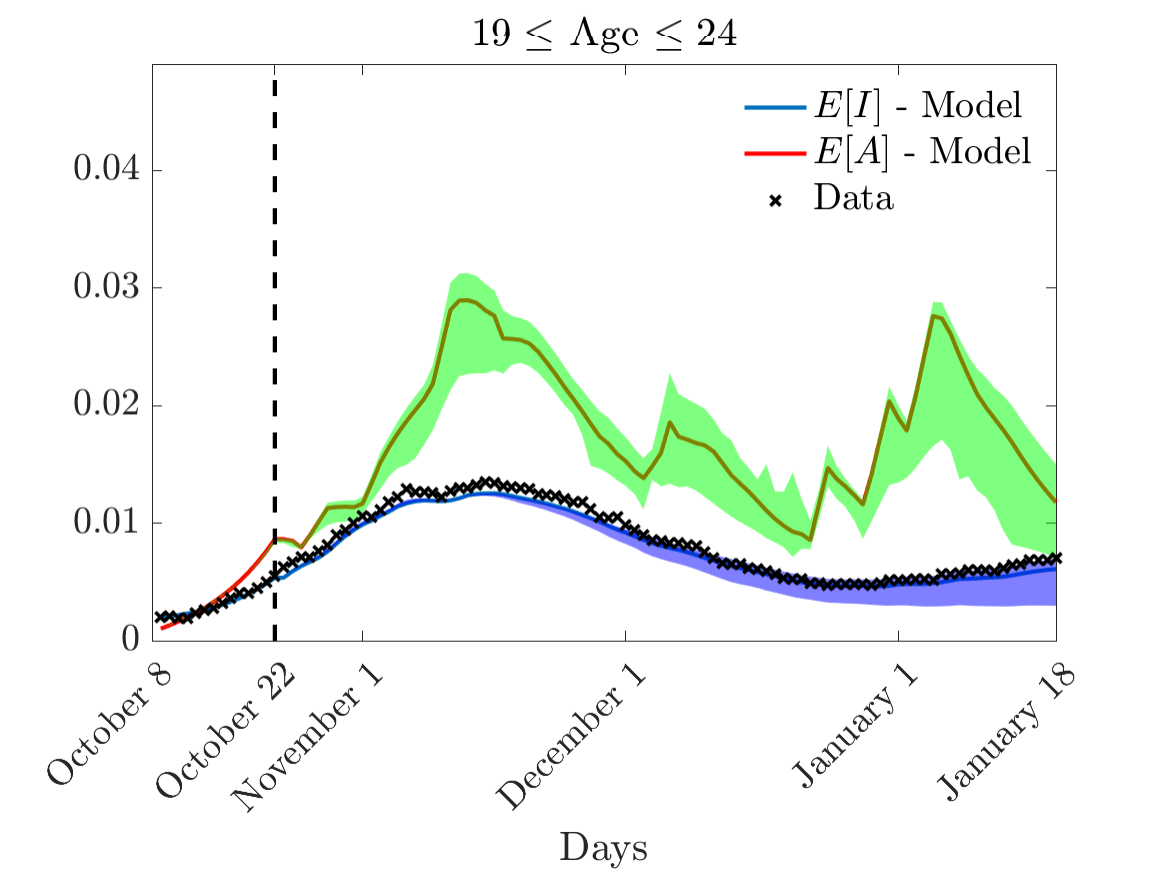}\\
\includegraphics[scale = 0.33]{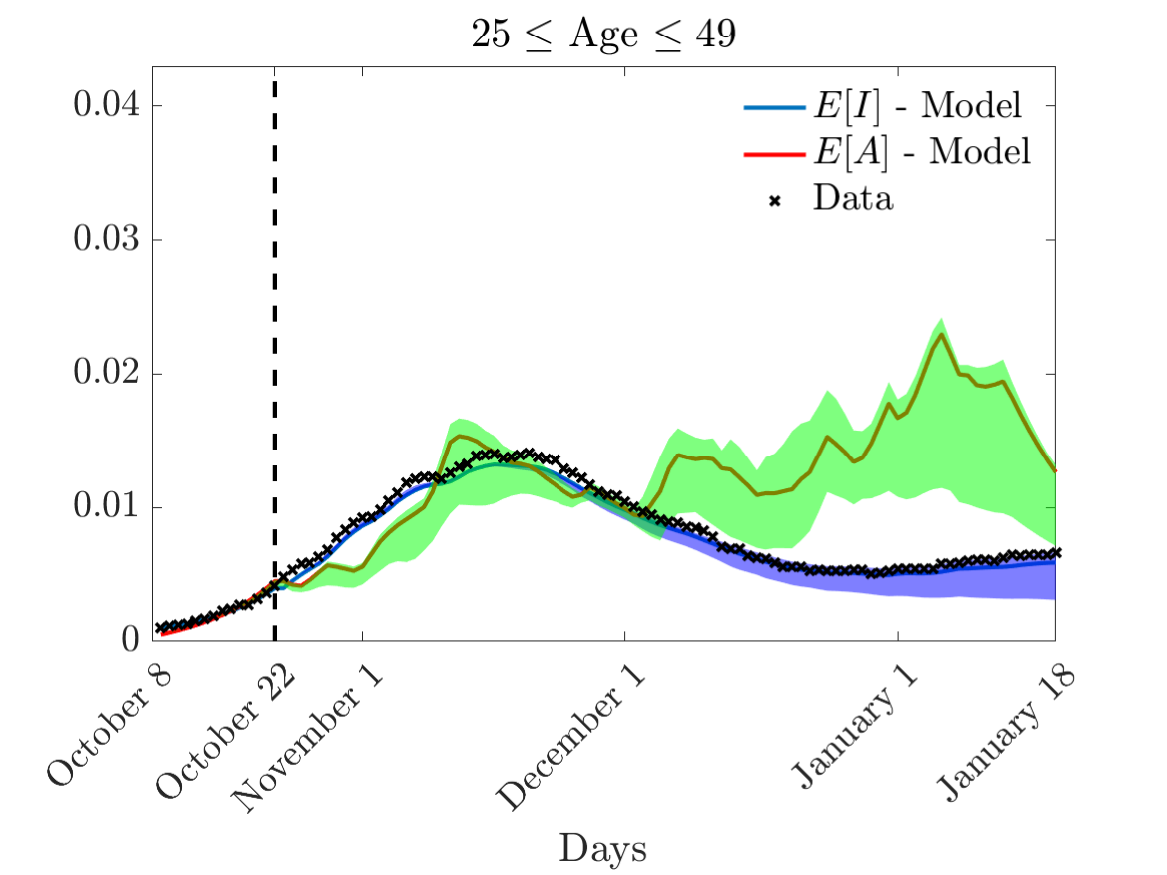}\hspace{-0.5cm}
\includegraphics[scale = 0.33]{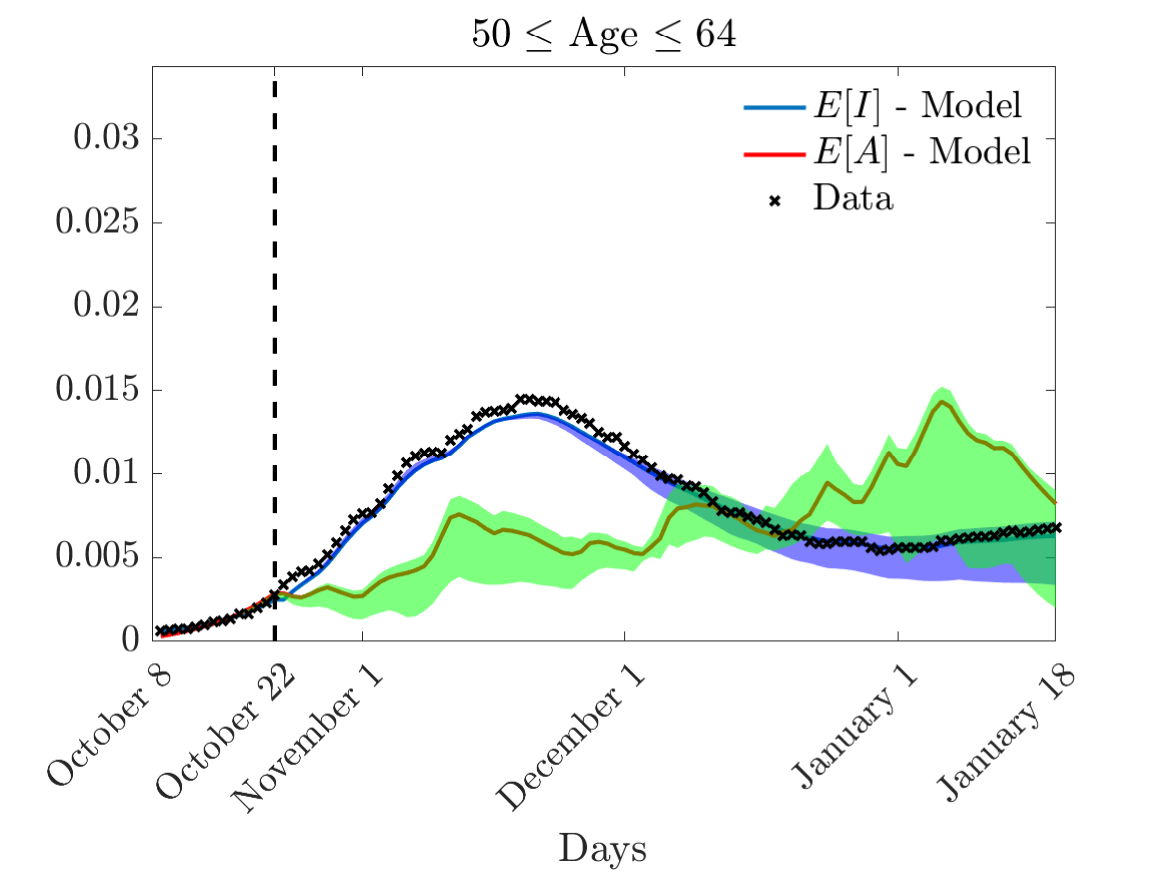}\\
\includegraphics[scale = 0.33]{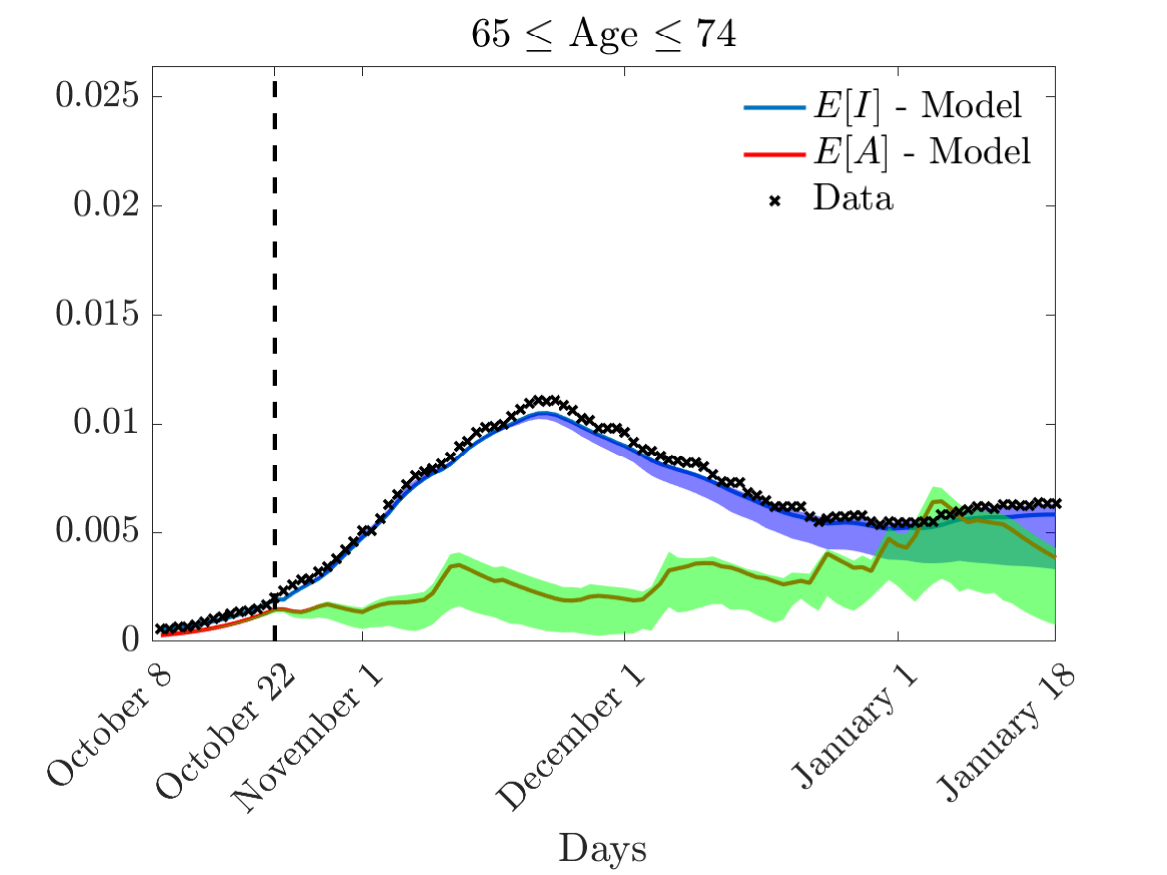}\hspace{-0.5cm}
\includegraphics[scale = 0.33]{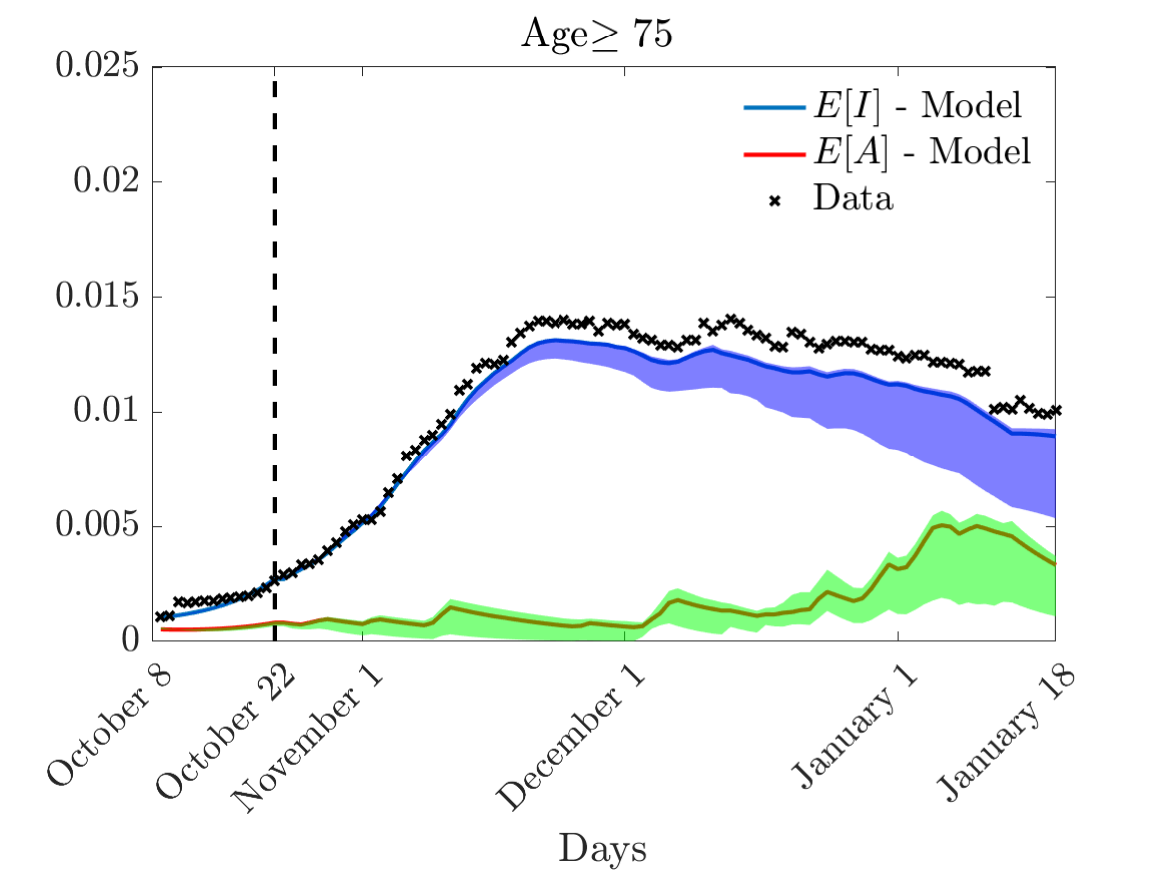}
\caption{Evolution of the expected number of infected and asymptomatic of the model \eqref{sir-closed_z} with uncertain recovery $\gamma_I(\z,x)$ and $\gamma_A(\z,x) = 2\gamma_I(\z,x)$ through the two-level optimization procedure and comparison with the data.}
\label{fig:cali}
\end{figure}

In Figure \ref{fig:cali}, we report the dynamics of confirmed cases with respect to the available data. Furthermore, we highlight in red the expected number of asymptomatic, i.e. $\mathbb E[\rho_A(\z,x,t)]$. Together with expected trends we plotted the confidence intervals obtained by imposing 
\[
\textrm{Prob}[A_\ell \le \rho_A(\z,x,t)\le A_r] = 1-c,
\]
with $c = 0.05$ for the $95\%$ confidence interval. The results show that the S-SIAR model is able to follow the experimental results in time for all age classes and then it can be used to forecast the epidemic trend and the possible effect of vaccination. It is worth to observe that the number of asymptomatic significantly changes in each age class decreasing as the age increases. This result naturally from the modeling choice and the optimization procedure and it is in agreement with the theoretical findings about the impact of the COVID-19 on the population. 

\begin{figure}
\centering
\includegraphics[scale = 0.4]{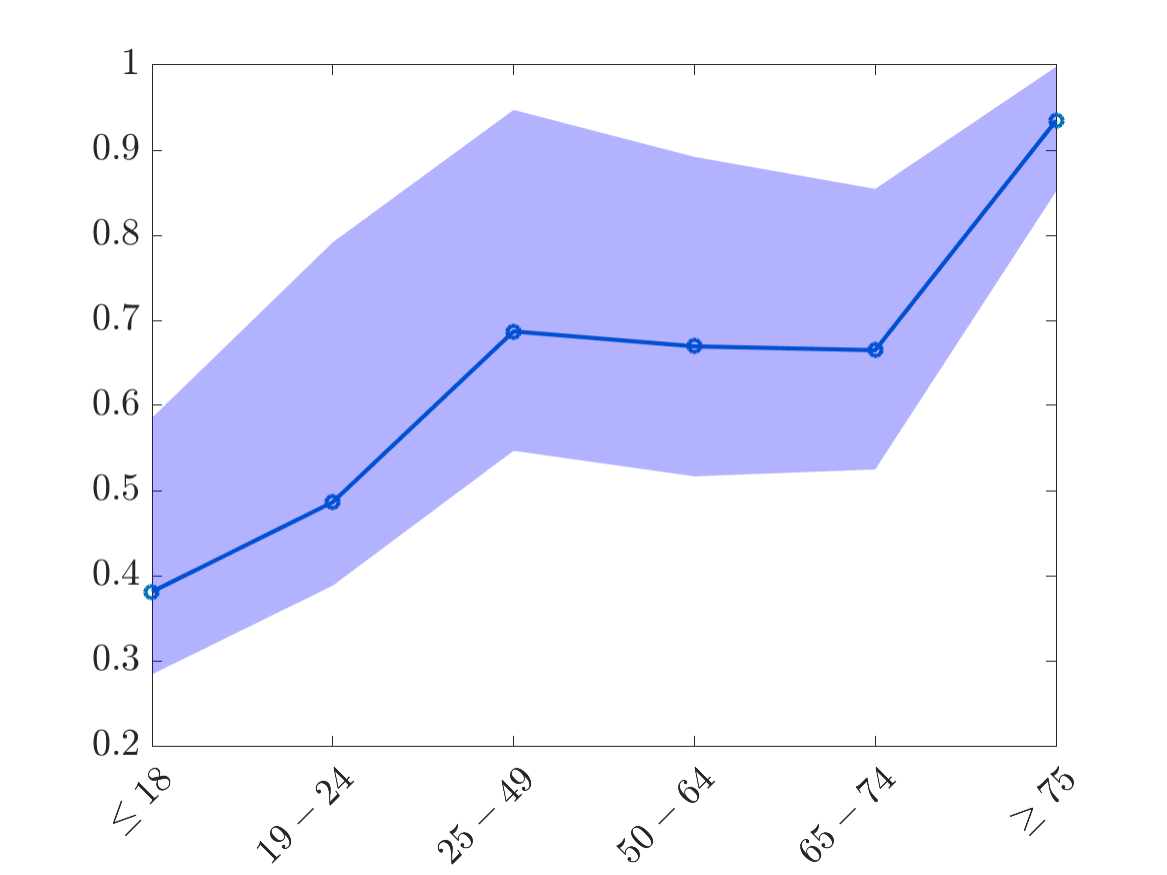}
\caption{Estimated contact reduction function $H_S$ by age over the reference time interval together with confidence bands. }
\label{fig:Hbar}
\end{figure}

In Figure \ref{fig:Hbar} we provide also the estimated function $H_S$ for the considered age-structure. As expected, the largest reduction of contacts is concentrated in the class with scholar activities. 

\subsection{Forecasting and vaccination strategies}\label{sec:forecasting}
In this part, we discuss some forecasting strategies using the model \eqref{sir-closed_z} with the parameters set in the previous part starting from our dataset of the Pavia province. The first simulation starts the 1st of January 2021 and takes as initial data the relative pandemic situation measured on that day. The second simulation starts one week after again using as initial condition the data relative then to the 7 of January 2021 in terms of susceptible, infected, asymptomatic and recovered individuals.
In order to use the proposed age-structured S-SIAR model, we introduce the recovery rates extrapolated from the data (see section \ref{sec:stat}) and we take into account the uncertainty in the determination of these values by introducing a bidimensional random space in which the two random variables are distributed as beta functions with parameters \eqref{tab:age_param} (see section \ref{sec:calibration}) as done in the previous section. Then, we consider the values $m_\beta(\z,x)$ (see equation \eqref{age-inci_z}) determined through the optimization procedure performed over the period of fall 2020 (between October 8 and October 22 2020 as explained again in the previous section). Finally, we take the average value of the contact functions $H_J(\z,x,t), \ J=S,I,A,R$ and the average number of asymptomatic $\xi(\z,x)$ over December 2020, i.e. the month before the first day of the first simulation for performing our short term forecast.

\begin{figure}
    \centering
    \includegraphics[scale = 0.43]{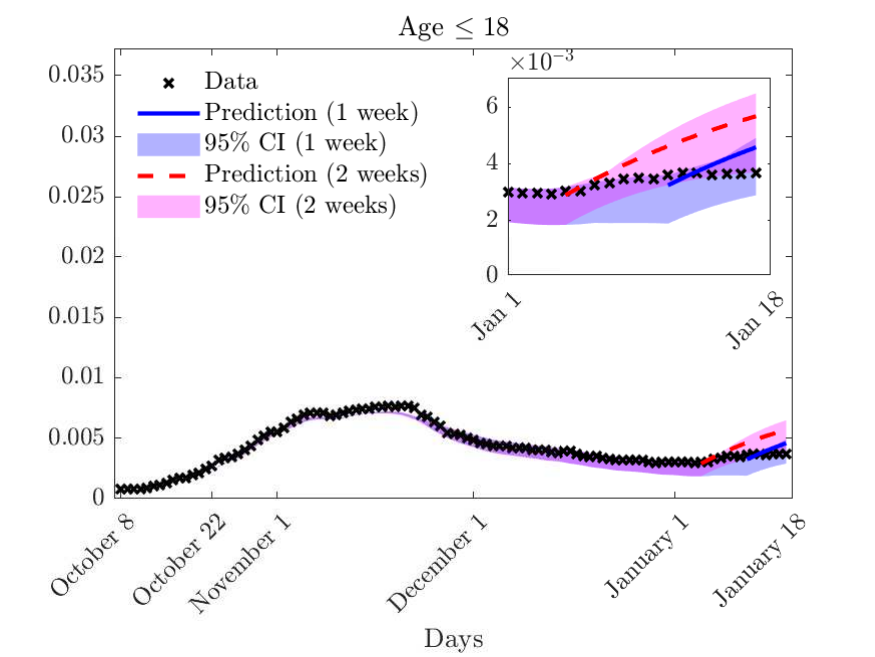}\hspace{-0.5cm}
    \includegraphics[scale = 0.43]{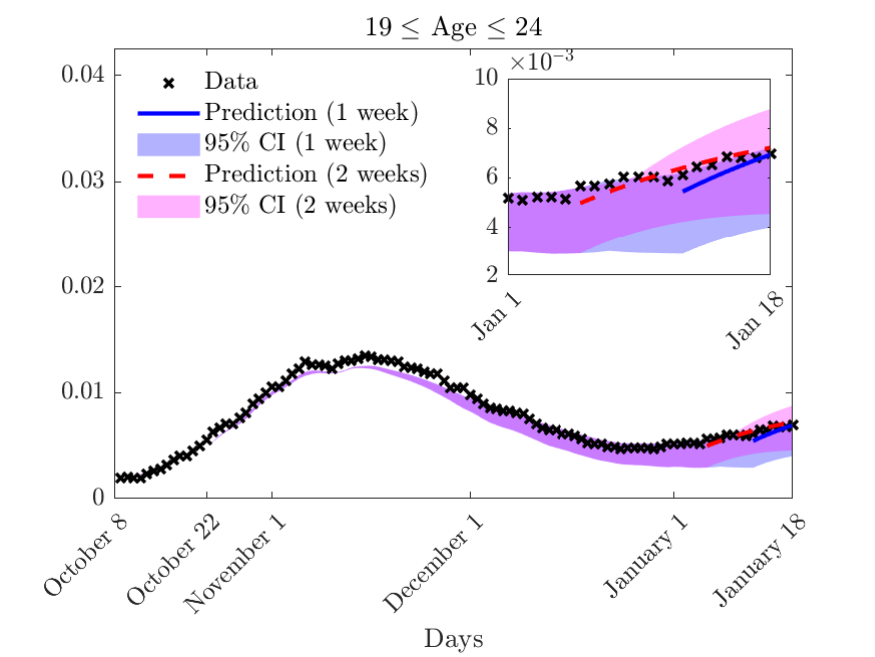}\\
    \includegraphics[scale = 0.43]{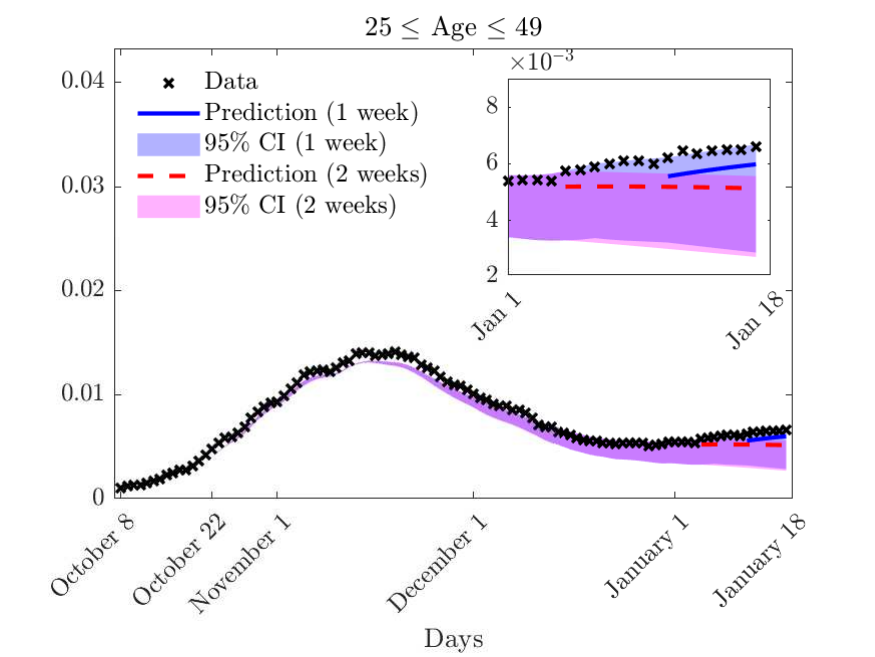}\hspace{-0.5cm}
    \includegraphics[scale = 0.43]{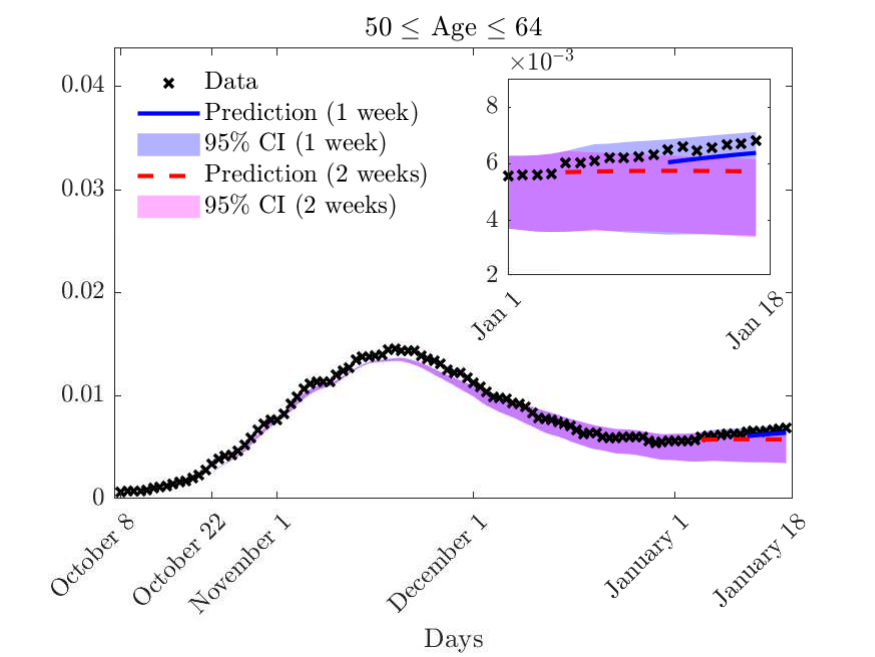}\\
    \includegraphics[scale = 0.43]{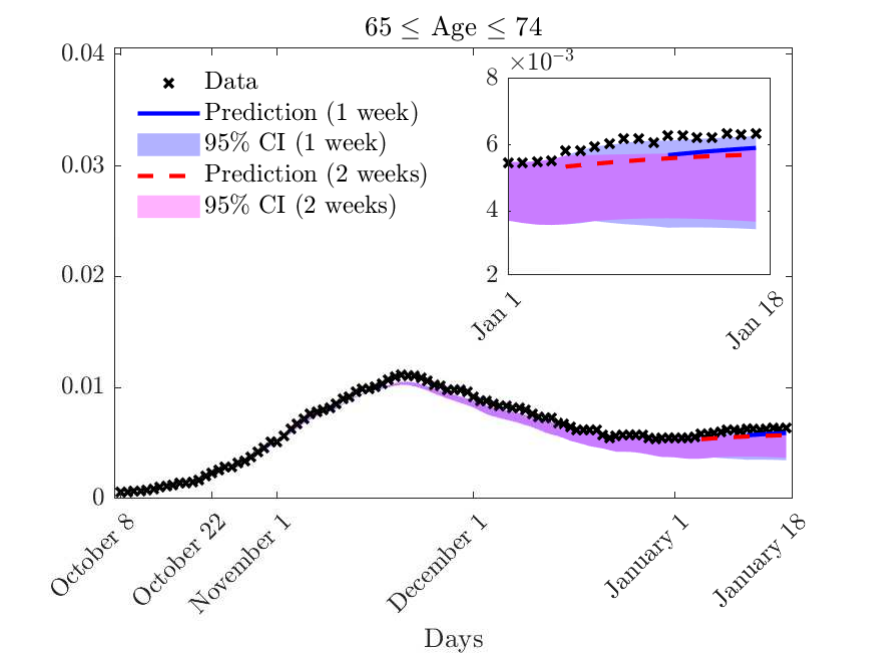}\hspace{-0.5cm}
    \includegraphics[scale = 0.43]{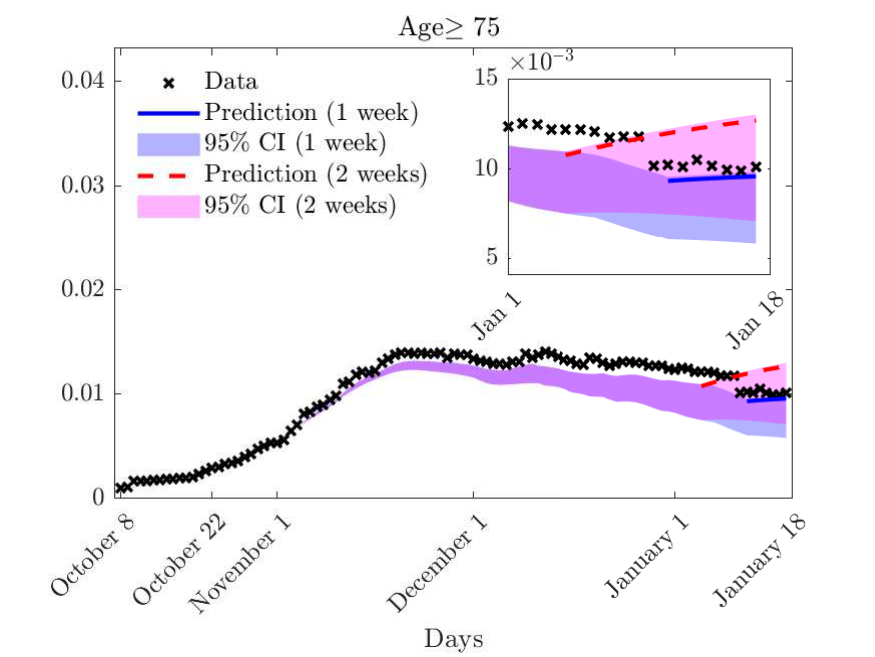}
    \caption{Comparison between 14 days and 7 days predictions based on S-SIAR model with age-structure.}
    \label{fig:prev}
\end{figure}

The Figure \ref{fig:prev} shows the results of the above detailed study in terms of number of infected classified by age. In particular, they show the comparisons between data and model for the two different described scenarios, i.e. one week prediction or two weeks prediction together with the confidence bands. The results are in very good agreement with observations.

\begin{figure}
    \centering
    \includegraphics[scale = 0.43]{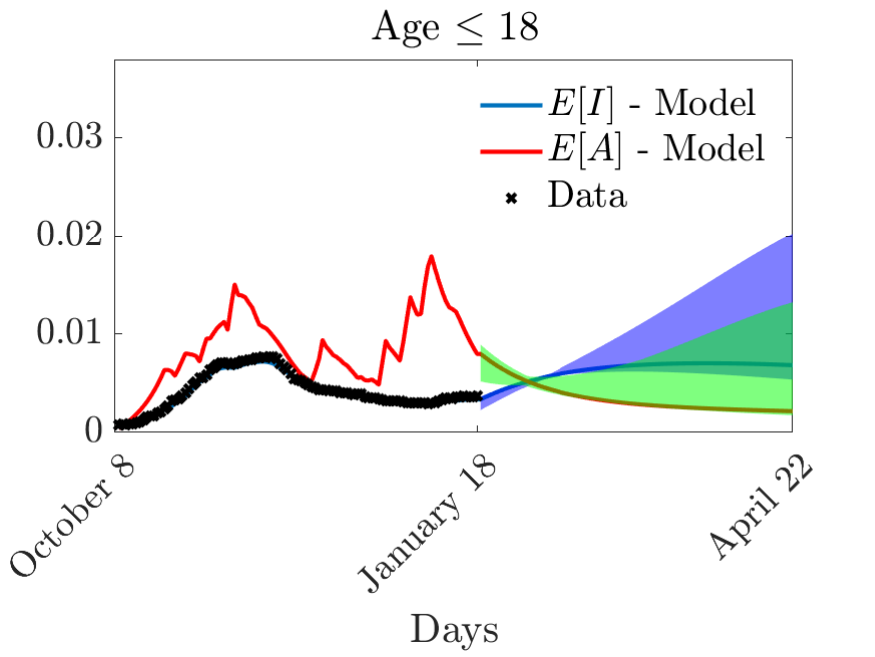}\hspace{-0.5cm}
    \includegraphics[scale = 0.43]{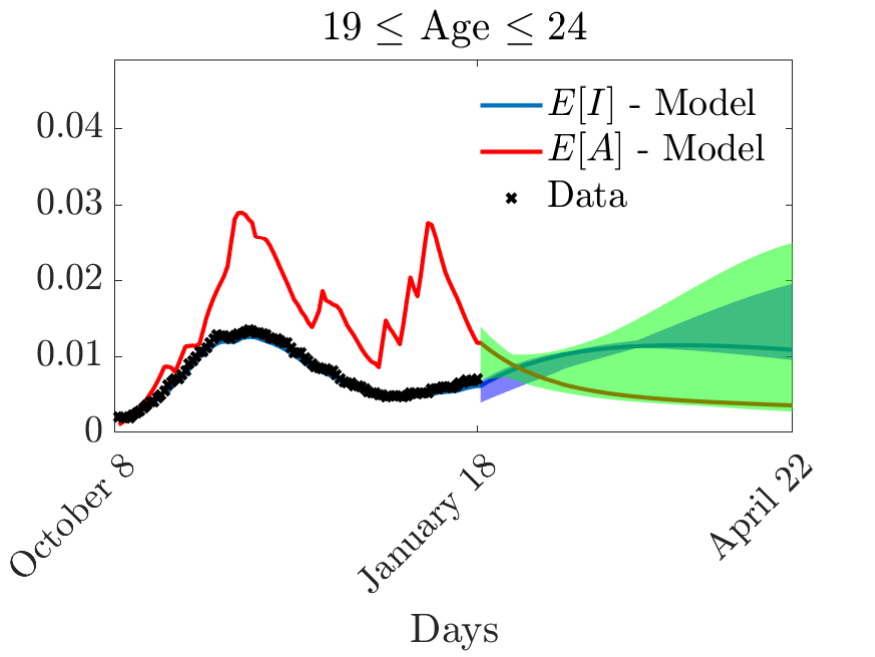}\\
    \includegraphics[scale = 0.43]{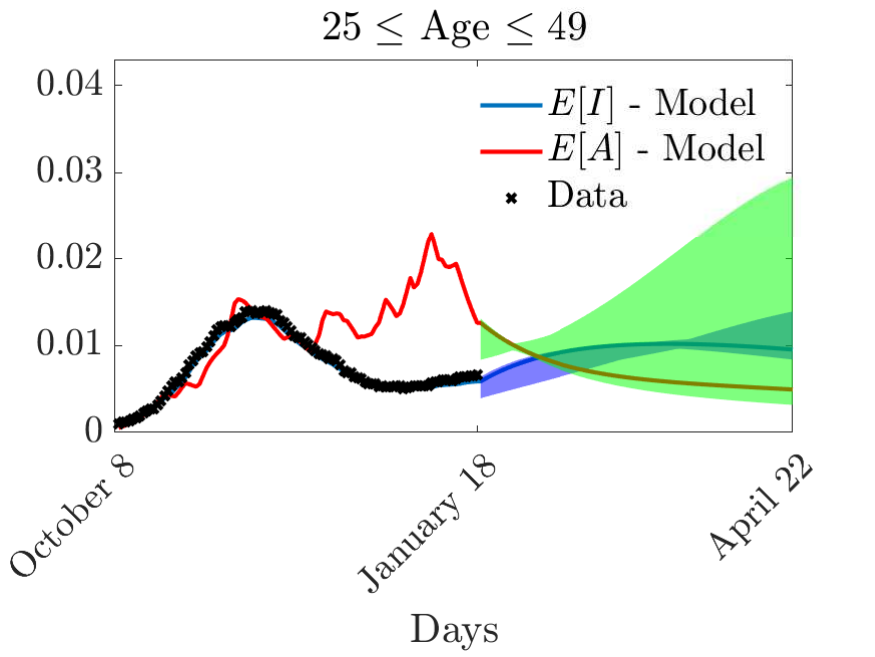}\hspace{-0.5cm}
    \includegraphics[scale = 0.43]{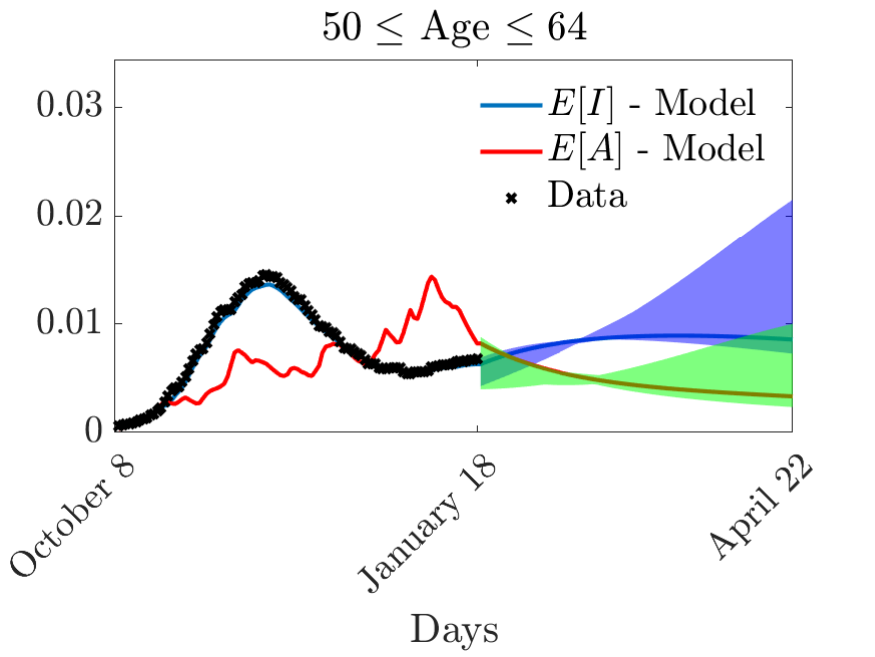}\\
    \includegraphics[scale = 0.43]{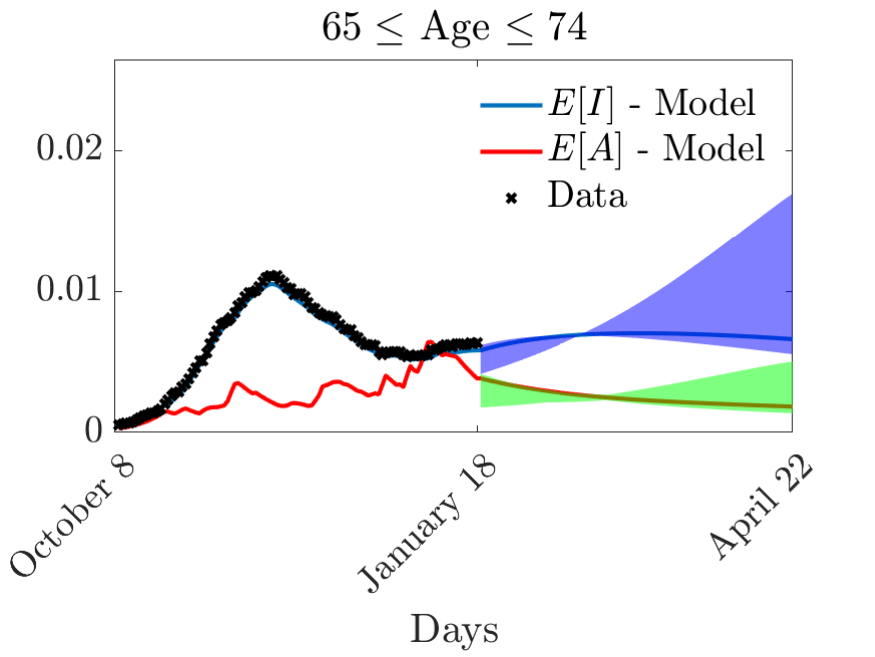}\hspace{-0.5cm}
    \includegraphics[scale = 0.43]{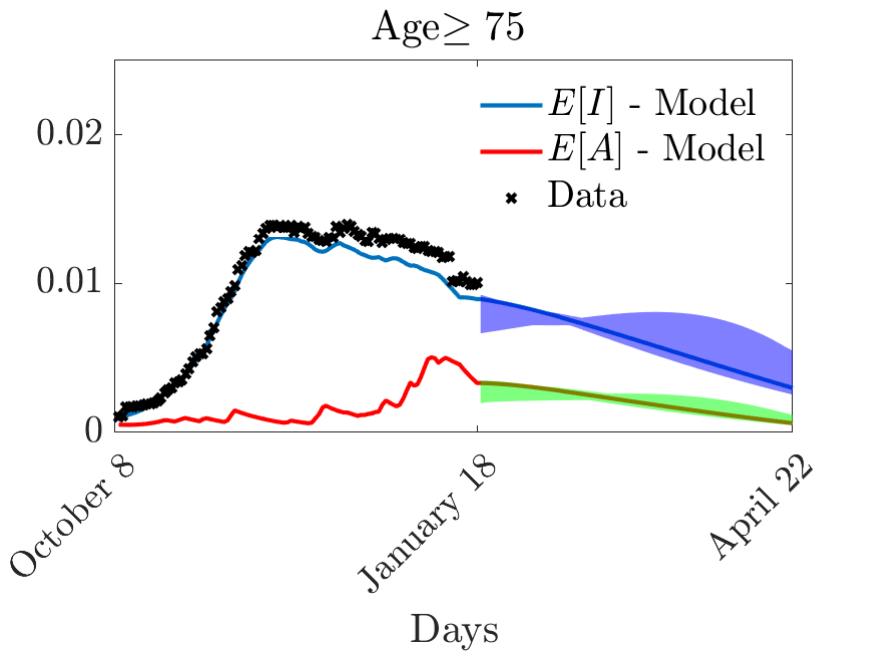}
    \caption{Simulated evolution over 90 days of the expected number of infected and asymptomatic using model \eqref{sir-closed_z} with age-priority vaccination campaign.}
    \label{fig:vacc1}
\end{figure}

\begin{figure}
    \centering
    \includegraphics[scale = 0.43]{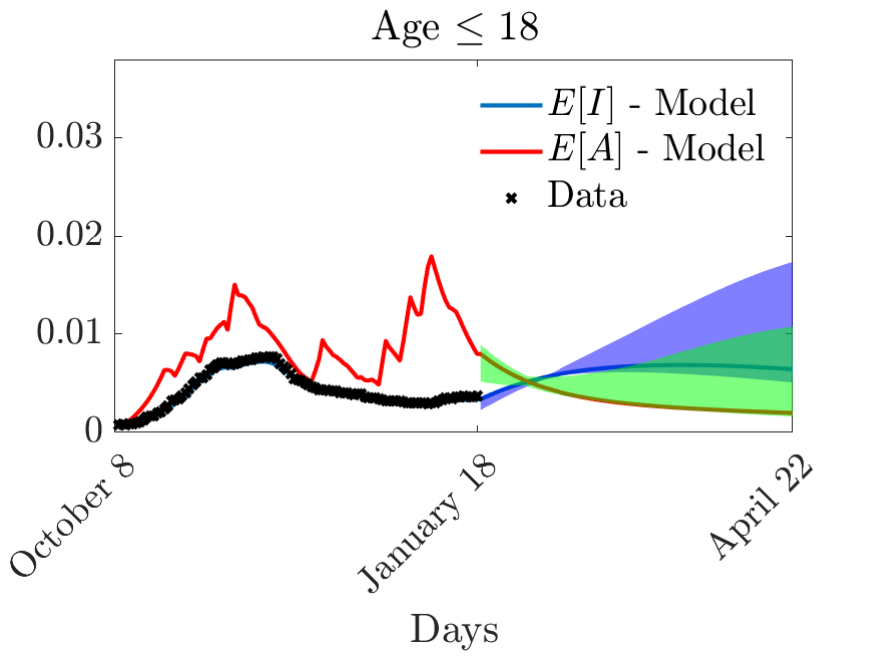}\hspace{-0.5cm}
    \includegraphics[scale = 0.43]{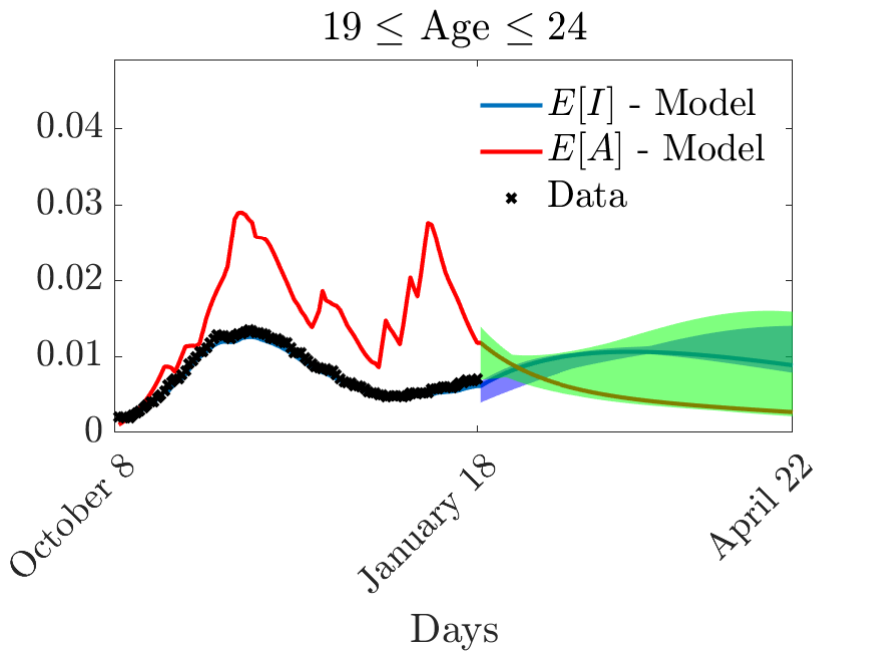}\\
    \includegraphics[scale = 0.43]{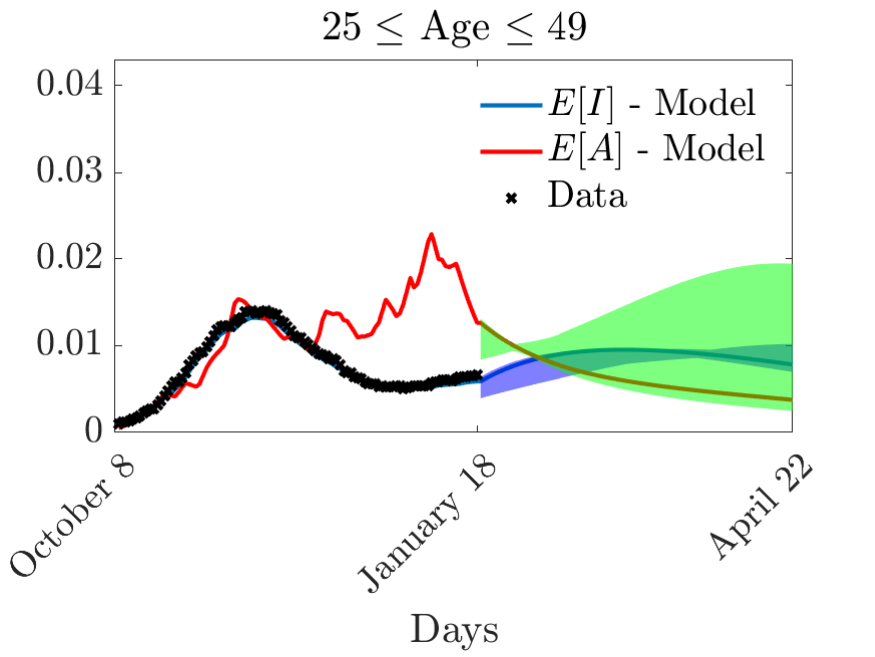}\hspace{-0.5cm}
    \includegraphics[scale = 0.43]{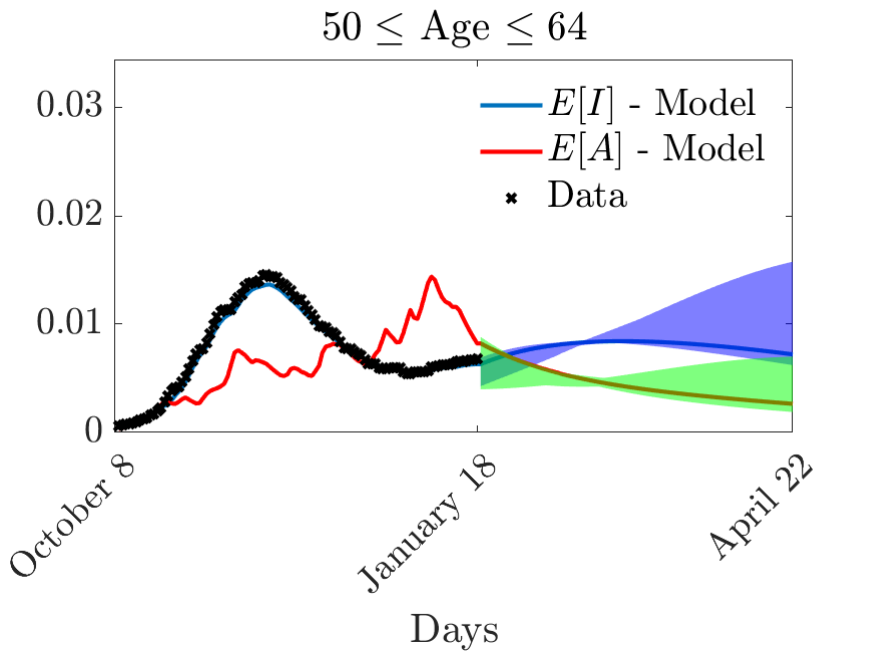}\\
    \includegraphics[scale = 0.43]{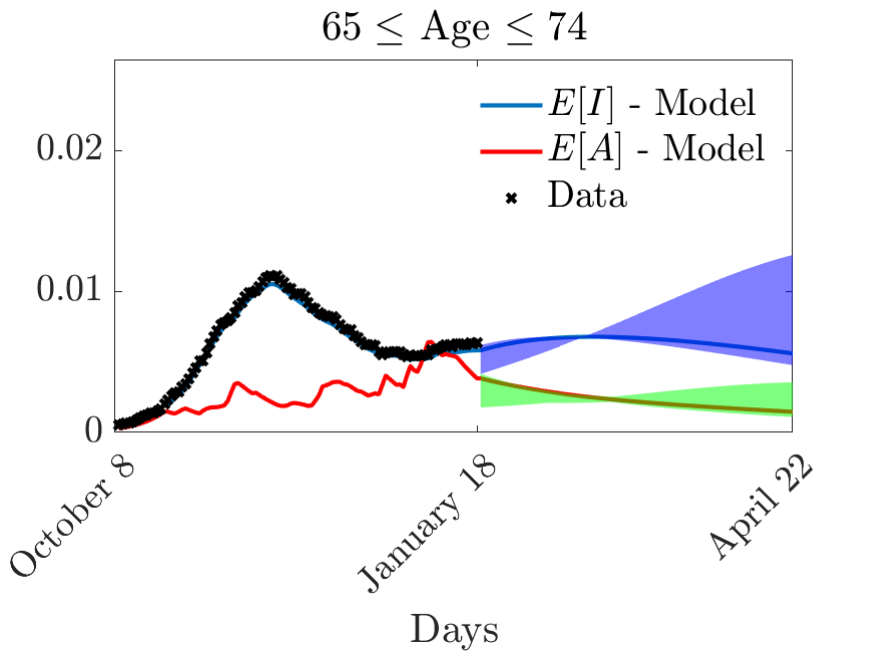}\hspace{-0.5cm}
    \includegraphics[scale = 0.43]{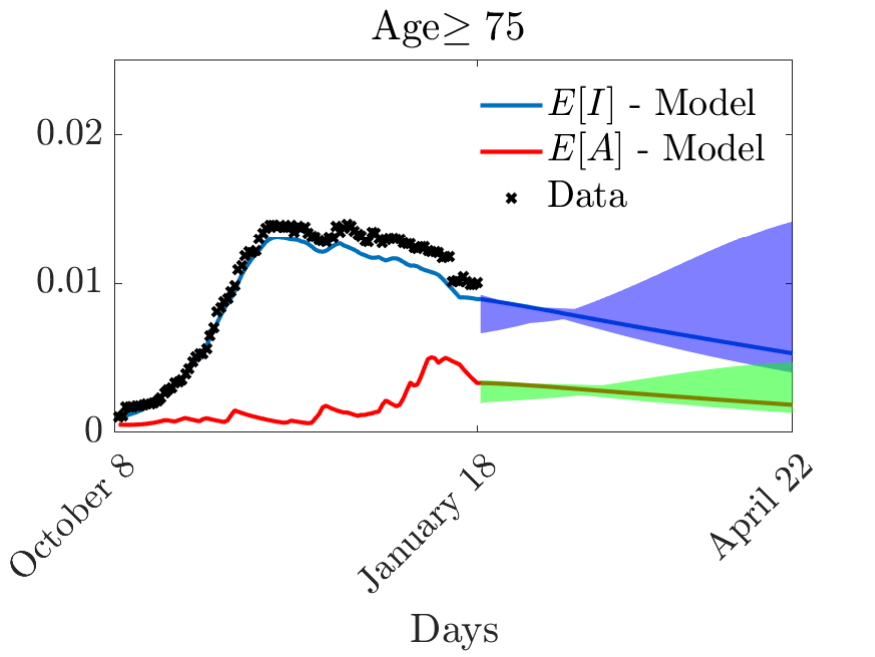}
    \caption{Evolution of the expected number of infected and asymptomatic using model \eqref{sir-closed_z} under an age uniform vaccination campaign.}
    \label{fig:vacc2}
\end{figure}

\begin{figure}
    \centering
    \includegraphics[scale = 0.43]{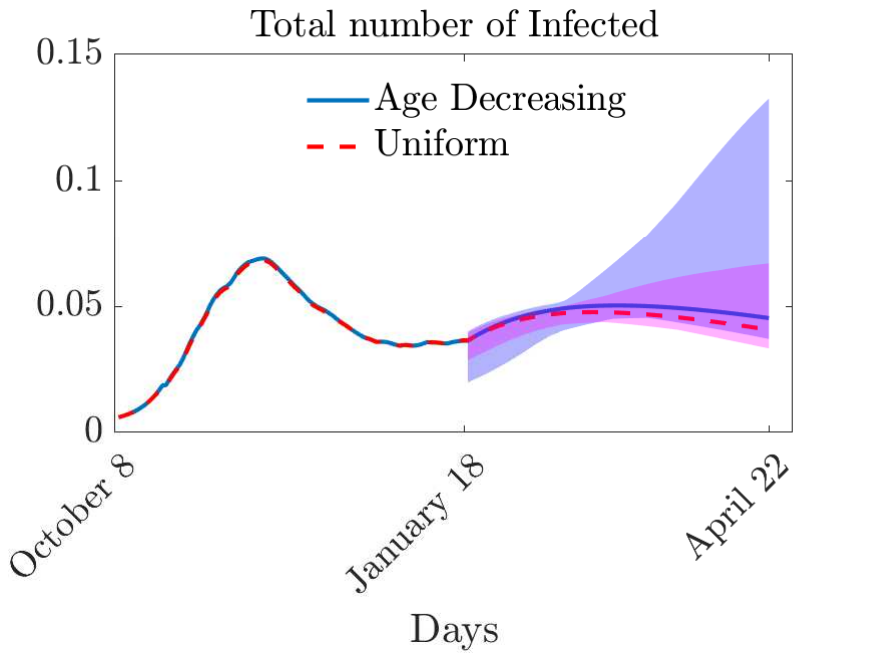}\hspace{-0.5cm}
    \includegraphics[scale = 0.43]{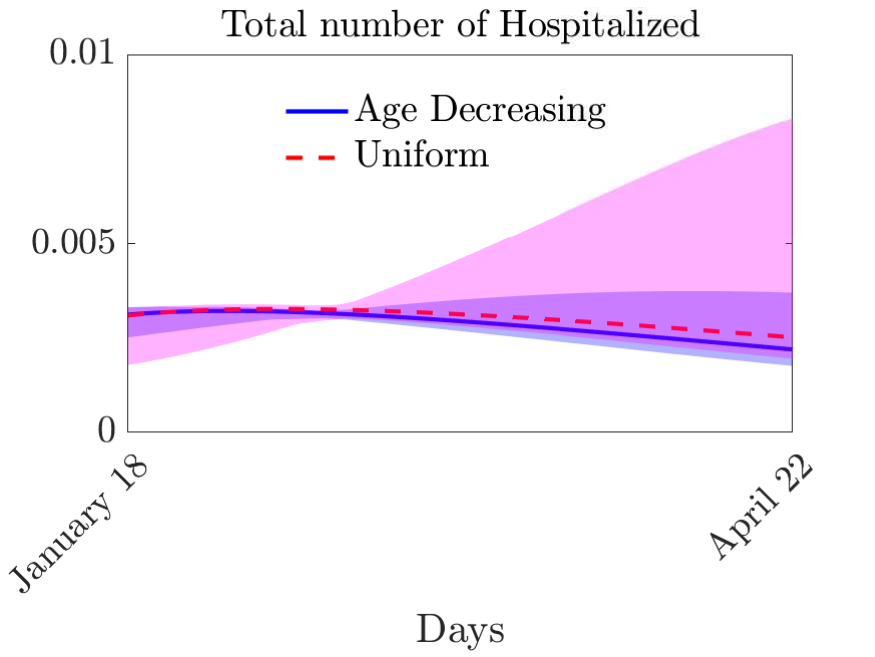}\\
    \caption{Comparison between uniform and age-priority vaccination campaign. Left: evolution of the total current number of infected. Right: expected total hospitalizations for the two studied vaccinations campaigns. }
    \label{fig:vacc3}
\end{figure}

As a last part, we discuss the role of the vaccination campaign. We consider two different possibilities. In the first case the vaccination campaign is implemented with an age-priority strategy, in the second case it is implemented uniformly over the population.
In both cases, we do the hypothesis that individuals belonging to the younger class do not vaccine themselves. We fix the number of $\bar\chi(t,x)$ per day as a consequence of the total number of vaccine at disposal. This means that we impose the following constraint
\be
\int_{\mathcal{I}}\bar\chi(t,x)dx=\bar\chi_{\textrm{max}}(t),
\ee 
where by hypothesis this quantity is such that every day the $1\%$ of the population have access to a vaccine. The first day in which the vaccination campaign starts corresponds to the last day of our dataset, i.e. the 18 of January and we perform a simulation over a period of 90 days. In Figure \ref{fig:vacc1} we report the results for the age-decreasing vaccination campaign while in Figure \ref{fig:vacc2} for the uniform vaccination campaign. The results show that as expected in the first case, the number of infected individuals over 75 decreases faster with respect to the second case. However, it is interesting to notice that the total number of infected is lower when a uniform vaccination is performed. The same holds true for the intensity of the pick of infected which is lower with the uniform campaign strategy as shown in Figure \ref{fig:vacc3} on the left. However, as it can be observed on the same figure on the right, the total number of hospitalized remains lower for the age-priority campaign as the larger number of hospitalized are older individuals. Therefore, in order to guarantee access to health system it appears convenient to introduce age-priority vaccinations. 

\section{Conclusions}\label{sec:conc}

\textcolor{black}{The development of strategies for mitigating the spreading of a pandemic is an important  public health priority. The recent case of COVID-19 pandemic has seen as main strategy for containing the spread restrictive measures on social contacts implemented through the so-called lockdown measures. This has been obtained by household quarantine, school or workplace closure, restrictions on travels, and, ultimately, on a total confining. Furthermore, to ensure efficacy containment policies it is of paramount importance to obtain robust prediction on their social effects through mathematical modelling. 
In this work, we introduced a system of kinetic equations coupling the emergence of the distribution of social contacts in Ref.  \cite{Plos} with the spreading of a pandemic driven by the rules of an age dependent SIAR model. The proposed modelling approach aims to explicitly quantify the pandemic transmission in terms of the number of social contacts of individuals. In particular, we measured through a data-driven approach the reduction of contacts in a pandemic scenario due to the non pharmaceutical interventions of local authorities and taking into account psychological effects in the contact dynamics. The proposed model has been demonstrated to be robust in terms of forecasting the trends of infection and hospitalization. Furthermore, the contact dynamics of asymptomatic population has been reconstructed to fit the available data of the province of Pavia (Lombardy region). Based on these considerations we have observed how an age decreasing vaccination strategy is capable to dampen hospitalizations. The results of such study represent a first step towards the use of such techniques by the local health authorities with the scope of furnishing a tool which permits to better organize the health facilities and the health workers in the Pavia region.}



\section{Appendix: Necessary facts from kinetic theory of rarefied gases and their relation with epidemic models} \label{sec:kinetic}

The Fokker-Planck system \fer{siar-gamma} contains all the information about the spreading of the epidemic in terms of the distribution of the classes as well as in terms of age and social contacts. Indeed, the knowledge of the densities $f_J(x,v,t)$, $J\in\{S,I,A,R\}$, allows to evaluate by integrations with respect to the $v$ variable all moments of interest. In particular, it permits to recover a set of equations describing the time evolution of susceptible, infected, asymptomatic and recovered individuals. However, in reason of the presence of the incidence rate $K(f_S,f_I+f_A)$, as given by \fer{inci}, the time evolution of the moments of the distribution functions is not explicitly computable, since, even in the simplest situation $\CU(x) = \CI$,  the evolution of a moment of a certain order depends on the knowledge of higher order moments, thus producing a hierarchy of equations which cannot be solved. Moreover, even if the knowledge of the solution of system \fer{siar-gamma} provides an exhaustive number of details, its computation in a reasonably extended interval of time would require a great effort from the numerical point of view. At the same time, the fitting of the parameters with such microscopic equations would be very difficult due to the lack of a sufficient number of experimental data. 

A relevant simplification of system \fer{siar-gamma} can be obtained by resorting to the following argument. As discussed in Ref.   \cite{DPeTZ}, the evolution of the pandemic is expected to be slower with respect to the so-called thermalization of the distribution of social contacts. This facts is expressed in system \fer{siar-gamma} by the value of the relaxation constant $\tau \ll 1$. Hence, we are in presence of a system in which one operator which acts mainly on the $x$ variable has a time scale of order $1$ (the SIR-type evolution of the densities among different classes) and this is the scale at which we are interested. On the other hand, the other operator characterizing the dynamics, acting mainly on the $v$ variable and describing the relaxation of the number of contacts, has a time scale $\tau \ll 1$. The presence of two times scales is a classical problem in kinetic theory, which is at the basis of the derivation of the equation of fluid dynamics from the Boltzmann equation \cite{Bob,Cer}.

We briefly survey the necessary results from classical kinetic theory that can be applied to the present situation and that allowed to obtain the system \eqref{sir-closed}. In order to maintain the analogies with the system \fer{siar-gamma}, the discussion that follows will be based on the kinetic Fokker--Planck equation \cite{Ris}. We mention however that the Fokker--Planck model is a consequence of a binary type of microscopic interactions which is typically described through Boltzmann type approach \cite{Cer}.
The Fokker--Planck equation is a fundamental model in  kinetic theories and statistical mechanics. It is a partial differential equation describing the time evolution of a density function $f(v,t)$, where $v\in \R^n, \, n \ge 1$ and $t \ge 0$, departing from a nonnegative initial density probability distribution $f_0(v)$. The standard assumptions on $f_0(v)$ is that it possesses finite mass $\rho$, mean velocity $u$ and temperature $\theta$, where for any given density $g(v)$
 \be\label{mass}
 \rho(g) = \int_{\R^n} g(v) \, dv
 \ee
is the mass density,
\be\label{velo}
 u(g) = \frac 1\rho \int_{\R^n}v g(v) \, dv
 \ee
is the mean velocity, and $\theta$ is the temperature defined by
 \be\label{temp}
 \theta(g)= \frac 1{n\rho} \int_{\R^n}|v-u|^2 g(v) \, dv.
 \ee
 The general form of the equation reads
 \be\label{FPgen}
\frac{\partial }{\partial t}f = Q(f) = \gamma \sum_{k=1}^n \left\{\frac{\partial^2 f}{\partial v_k^2} + \frac 1{\theta(f)}  \frac{\partial }{\partial v_k}[(v_k-u_k(f))f]\right\}.
\ee
The one-particle friction constant $\gamma$ is usually assumed to be a function of $\rho, u, \theta$. Equation \fer{FPgen} has a stationary solution of given mass $\rho$, mean velocity $u$ and temperature $\theta$ given by the Maxwellian density function
 \be\label{Max}
 M_{\rho,u,\theta}(v) = \rho \, \frac 1{(2\pi\theta)^{n/2}}\exp\left\{-\frac{|v-u|^2}{2\theta}\right\},
 \ee
which belongs to the kernel of the operator $Q$, so that $Q(M_{\rho,u,\theta})=0$. Note that, in view of its differential structure, mass density, mean velocity and temperature are preserved in time by the Fokker-Planck equation \fer{FPgen}.

In view of its conservations and its decay property towards the equilibrium Maxwellian distribution,\cite{AMTU} the Fokker-Planck operator can be fruitfully used in place of the Boltzmann collision operator,\cite{Cer} to describe the evolution of the rarefied gas phase space density  $f(x,v,t)$, with $x,v \in \R^3$ 
 \be\label{bol1}
 {\partial\over \partial t}f(x,v,t) = A(f)(x,v,t)  +
 \frac 1\varepsilon Q(f(x,v,t)).
 \ee
 This equation contains terms accounting for the two ways that the  density can change. The operator
 $$A(f)(x,v,t) = -v\cdot\nabla_x f(x,v,t)$$  represents the effects of {\it  transport};
 that is, the motion
 \be\label{tran}
 x_0 \mapsto x_0 + (t-t_0)v_0\qquad v_0 \mapsto v_0
 \ee
 of molecules between interactions. The Fokker--Planck operator $Q(f)$  represents the effects
 of interactions and describes relaxation to the local Maxwellian equilibrium,\cite{Cer,Cer94} as a function of the local  mass $\rho(x,t)$, velocity $u(x,t)$
and temperature $\theta(x,t)$:
 \begin{equation} \label{equi-M}
M(x,v,t) = \frac{\rho(x,t)}{(2 \pi \theta(x,t))^{3/2}} \exp\left \{ -
\frac{|v- u(x,t)|^2}{2 \theta(x,t)} \right \}.
\end{equation}
Last, $\varepsilon$ in \fer{bol1} is a suitable relaxation time which plays the same role of the scaling parameter $\tau$ in the epidemic model \eqref{siar-gamma}. 

The analogy which links collisional kinetic theory to social problems relies on the natural assumption that a system composed of a sufficiently large number of agents can be described using the laws of statistical mechanics \cite{PT13}. Indeed, in the proposed model \fer{siar-gamma}, there is an almost literal translation of concepts: the \emph{molecules} of a gas are identified with the
\emph{individuals}, the \emph{energy of particles} correspond to the \emph{number of contacts of individuals}, and relaxation consequent to \emph{collisions} translate into relaxation consequent to \emph{modification of social contacts}. 

Further, consider that we can write each equation of system \fer{siar-gamma} 
as 
\be\label{bol2}
{\partial\over \partial t}f_J(x,v,t) = A_J ({\bf f})(x,v,t)  +
 \frac 1\varepsilon Q_J(f_J(x,v,t)), \quad J \in \{S,I,A,R\},
\ee
where the operators $A_J$ describe the evolution of the epidemic system, and ${\bf f} = (f_S,f_I,f_A,f_R)$ is a vector.  Both the equations \fer{bol1} and \fer{bol2} contain two distinct operators which play different roles. The  \emph{transport} operator $A(f)$ in \fer{bol1} corresponds to the operators $A_J$  in \fer{bol2}. Also, the Fokker--Planck operator in \fer{bol1} describing relaxation to the local Maxwellian equilibrium \fer{equi-M} corresponds to the Fokker--Planck type operators $Q_J$, $J \in \{S,I,A,R\}$ in \fer{bol2} describing relaxation to the equilibrium density of social contacts.

Resorting to the proposed analogies between individuals and colliding particles, various well established methods from kinetic theory and statistical physics are ready for application to system \fer{siar-gamma}. Most notably, these numerous tools can now be used to analyze the evolution of the epidemic. In this way, the kinetic description of social contacts provides naturally one possible explanation for the development of the universal profiles observed in these situations.

A non secondary aspect of this analogy is the possibility to resort to the classical closure procedure around the local equilibrium density to recover the underlying equations of fluid dynamics. That is what we did by passing from \eqref{siar-gamma} to \eqref{sir-closed}. Indeed, the local Maxwellian equilibrium density \fer{equi-M} allows to obtain all moments of the distribution in terms of the principal ones, given by mass density, mean velocity and temperature. While in the classical kinetic theory, the particle density depends on the space variable $x$, the velocity
variable $v$ and time $t$, in the framework of epidemic one can study the evolution of the distribution functions of the individuals which depends on age $x
\in \CI$, social contacts $v \in \R_+$ and time $t \in \R_+$, denoted by
$f_J=f_J(x,v,t)$, $J\in\{S,I,A,R\}$. By continuing to follow the analogy with kinetic theory, it is useful to emphasize the role of the different
parameters by identifying the \textit{velocity} with the \textit{social contacts}, and the
\textit{position} with the \textit{age}. By doing this, one assumes at once that the variation of the distributions $f_J(x,w,t)$ with respect to the social contact parameter $v$ will depend on the \emph{relaxation} term (the Fokker--Planck dynamics), while the change of distributions in terms of the age $x$ depends on the SIAR \emph{transport} term, which contains
the equation of motion, namely the law of variation of $x$ with
respect to time due to the passage from one class to the other. 

Thanks to the above considerations, it appears natural to assume that the  Fokker--Planck equations
 \[
\frac{\partial f_J(v,t)}{\partial t} = Q_J(f_J)(v,t)=  \frac{\partial}{\partial v}\left[ \nu \frac{\partial}{\partial v} (v f_J(v,t)) +  \left(\frac v{\bar m_J} -1\right)f_J(x,v,t)\right],
\]
$J\in \{S,I,A,R\}$, play the same role as the Fokker--Planck equation \fer{FPgen} in equation \fer{bol1}, and consequently describe relaxation of the distribution of the social contacts to the local Gamma-type equilibria with local masses $\rho_J(x,t)$, and local mean densities of contacts $\bar m_J(x,t)$, as defined by \fer{densi} and, respectively \fer{mean-number}
\be\label{gamma-JJ}
M_J^\infty(x,v,t) =  \rho_J(x,t) \left(\frac \nu{\bar m_J(x,t)}\right)^\nu \frac 1{\Gamma\left(\nu \right)} v^{\nu -1}
\exp\left\{ -\frac\nu{\bar m_J(x,t)}\, v\right\}.
 \ee 
Finally, a direct and clear understanding of the derivation of macroscopic equations relies on the so-called fractional step method consisting in considering separately and sequentially, in each (small) time step, the
transport and relaxation operators. In consequence, during this short time interval one recovers the
evolution of the density $f$, solution to \fer{bol1} from the joint action of the relaxation
\begin{equation}\label{id-rel}
 {\partial\over \partial t}f(x,v,t) = 
 \frac 1\varepsilon Q(f(x,v,t)).
\end{equation}
and transport
\begin{equation}\label{id-tra}
 {\partial\over \partial t}f(x,v,t) = A(f)(x,v,t).
\end{equation}
Since mass, mean velocity and temperature are conserved by the Fokker--Planck operator, the conservation's properties (for fixed $x$), are enough to guarantee that \fer{id-rel} pushes the solution towards the (local $x$-dependent) locally Maxwellian equilibrium \fer{equi-M}. Then, if $\varepsilon$ is sufficiently small, one can easily argue that the solution to \fer{id-rel} is \emph{sufficiently
close} to the equilibrium \fer{equi-M}, and this equilibrium can be used in the
transport step \fer{id-tra} to find closed equations for the local mass, mean velocity and temperature. These equations are nothing but the Euler equations of fluid dynamics \cite{Cer}.  In analogous way, we can proceed to close the system \fer{siar-gamma} around the local equilibrium \fer{gamma-JJ} and this gives finally the equations \eqref{sir-closed} used in the paper.

\section*{Acknowledgement} This work has been written within the activities of the Agreement 60/2020, Protocol number 154768 of December 21, 2020, between the Department of Political and Social Sciences of the University of Pavia and the Health Protection Agency (ATS) of the province area of Pavia, with object ``Mathematical modeling and statistics for the forecast of the COVID-19 epidemic in the territory of the Province of Pavia'', and within the activities  of GNFM group  of INdAM (National Institute of
High Mathematics). The research has been partially supported  by
the Italian Ministry of Education, University and Research (MIUR): Dipartimenti
di Eccellenza Program (2018--2022) - Dept. of Mathematics "F.
Casorati", University of Pavia, and by the PRIN Project 2017 (2017TEXA3H) entitled ``Optimal mass
transportation, geometrical and functional inequalities with applications''.  \\
G.D. would like to thank the Italian Ministry of Instruction, University and Research (MIUR) to support this research with funds coming from PRIN Project 2017 (No.2017KKJP4X) entitled  ``Innovative numerical methods for evolutionary partial differential equations and applications''.

\end{document}